\documentclass[preprint]{aastex63}

\usepackage{graphicx}
\usepackage[FIGTOPCAP]{subfigure}
\usepackage{amsmath,booktabs,threeparttable,url, bm}
\usepackage[hyphenbreaks]{breakurl}
\usepackage{CJKutf8}
\usepackage{xspace}

\newcommand{\Msun}{\ensuremath{\mathrm{M}_\odot}\xspace}

\newcommand{\cntext}[1]{\begin{CJK*}{UTF8}{bsmi}#1\end{CJK*}}

\usepackage{natbib}
\received{\today}
\submitjournal{ApJ}

\shorttitle{BH formation and MMA}
\shortauthors{Pan et al.}

\begin{document}
\title{Stellar Mass Black Hole Formation and Multi-messenger Signals from Three Dimensional Rotating Core-Collapse Supernova Simulations}

\newcommand*{\NTHUP}{Department of Physics, National Tsing Hua University, Hsinchu 30013, Taiwan}
\newcommand*{\NTHUA}{Institute of Astronomy, National Tsing Hua University, Hsinchu 30013, Taiwan}
\newcommand*{\CICA}{Center for Informatics and Computation in Astronomy, National Tsing Hua University, Hsinchu 30013, Taiwan}
\newcommand*{\BASEL}{Departement Physik, Universit\"{a}t Basel, Klingelbergstrasse 82, CH-4056 Basel, Switzerland} 
\newcommand*{\PAMSU}{Department of Physics and Astronomy, Michigan State University, East Lansing, MI 48824, USA}
\newcommand*{\CMSE}{Department of Computational Mathematics, Science, and Engineering, Michigan State University, East Lansing, MI 48824, USA}
\newcommand*{\NSCL}{National Superconducting Cyclotron Laboratory, Michigan State University, East Lansing, MI 48824, USA}
\newcommand*{\JINA}{Joint Institute for Nuclear Astrophysics-Center for the Evolution of the Elements, Michigan State University, East Lansing, MI 48824, USA}
\newcommand*{\GSI}{GSI Helmholtz Center for Heavy Ion Research, D-64291 Darmstadt, Germany}


\author[0000-0002-1473-9880]{Kuo-Chuan Pan (\cntext{潘國全})}
\affiliation{\NTHUP} \affiliation{\NTHUA} \affiliation{\CICA}  

\author{Matthias Liebend\"{o}rfer}
\affiliation{\BASEL}

\author[0000-0002-5080-5996]{Sean~M.~Couch}
\affiliation{\PAMSU} \affiliation{\CMSE} \affiliation{\NSCL} \affiliation{\JINA}

\author[0000-0002-7256-9330]{Friedrich-Karl Thielemann}
\affiliation{\BASEL} \affiliation{\GSI}
 

\begin{abstract}

We present self-consistent, 3D core-collapse supernova simulations of a 40-$\Msun$ progenitor model
using the isotropic diffusion source approximation for neutrino transport and an effective general 
relativistic potential up to $\sim0.9$~s~postbounce. 
We consider three different rotational speeds with initial angular velocities of $\Omega_0=0$,~0.5, and~1~rad~s$^{-1}$ 
and investigate the impact of rotation on shock dynamics, black hole formation, and gravitational wave signals. 
The rapidly-rotating model undergoes an early explosion at $\sim 250$~ms postbounce and shows signs of the low $T/|W|$ instability. 
We do not find black hole formation in this model within $\sim 460$~ms postbounce.
In contrast, we find black hole formation at 776~ms~postbounce and 936~ms~postbounce for the non-rotating 
and slowly-rotating models, respectively. 
The slowly-rotating model explodes at $\sim 650$~ms postbounce, and the subsequent fallback accretion onto the proto-neutron star (PNS) results in BH formation. 
In addition, the standing~accretion~shock~instability induces rotation 
of the proto-neutron star in the model that started with a non-rotating progenitor. 
Assuming conservation of specific angular momentum during black hole formation, 
this corresponds to a black~hole spin parameter of $a=J/M=0.046$. 
However, if no explosion sets in, all the angular momentum will eventually be accreted by the BH, resulting in a non-spinning BH.
The successful explosion of the slowly-rotating model drastically slows down the accretion onto the PNS, allowing continued cooling and contraction that results in an extremely high gravitational-wave frequency ($f\sim3000$~Hz) at black~hole formation,   
while the non-rotating model generates gravitational wave signals similar to our corresponding 2D simulations.  

\end{abstract}

\keywords{Core-collapse supernovae (304); Black holes (162); Neutron stars (1108); Gravitational wave astronomy (675); Hydrodynamical simulations (767)} 



\section{INTRODUCTION}

Detection of gravitational waves (GWs) from a nearby core-collapse supernova (CCSN) 
could be the next milestone of GW astronomy and multimessenger astrophysics.  
Improving GW search pipelines by providing a comprehensive understanding of CCSN gravitational waveforms derived from numerical simulations could be {\it crucial} to making such a detection a reality \citep{2016PhRvD..94j2001A, 2020PhRvD.101h4002A} 
and to design the next generation detectors \citep{2019PhRvD..99f3018R}.    
In the past decade, our understanding of CCSN explosion engine(s) has advanced dramatically due to advancing high-performance computing facilities \citep{2017hsn..book.1095J}, allowing us to perform high-fidelity CCSN simulations with detailed microphysics and sophisticated neutrino transport 
\citep{2015ApJ...807L..31L, 2016ApJ...829L..14K, 2018ApJ...855L...3O,
2018MNRAS.477L..80K, 2018ApJ...852...28S, 2018ApJ...852L..19C, 2019MNRAS.486.2238A, 2018ApJ...865...81O, 2018arXiv180101914M, 2019MNRAS.486.2238A, 2019ApJ...876L...9R, 2020MNRAS.491.2715B}. 
Furthermore, the community has reached some consistency and agreement in spherically symmetric simulations \citep{2018JPhG...45j4001O} 
and well-controlled multi-dimensional simulations \citep{2018A&A...619A.118C, 2019JPhG...46a4001P, 2019ApJ...873...45G}.
However, the parameter space of CCSN explosion models and multimessenger signal predictions from multi-dimensional simulations is not yet fully explored.    

Previous studies include the dependence on the progenitor mass \citep{2020MNRAS.491.2715B}, nuclear Equation of State (EoS) 
\citep{2016ApJ...829L..14K, 2018ApJ...857...13P, 2020arXiv200110434S}, 
grid resolution \citep{2015ApJ...808...70A, 2019MNRAS.490.4622N, 2019arXiv190401699M},
rotational effects \citep{2016MNRAS.461L.112T, 2018ApJ...852...28S, 2019MNRAS.486.2238A, 2020MNRAS.tmp.1245P, 2020MNRAS.tmpL..27S}, 
or magneto-hydrodynamical effects \citep{2014ApJ...785L..29M, 2020MNRAS.492.4613O}.
Those studies suggest that CCSNe with non-rotating intermediate-mass progenitors 
show weak GW emissions, which can be observed only for galactic CCSNe 
\citep{2016ApJ...829L..14K, 2018ApJ...865...81O, 2018arXiv180101914M, 2019MNRAS.486.2238A, 2019ApJ...876L...9R}. 
Although the likelihood of a CCSN occurring within the Milky Way is small, extreme conditions, 
such as massive progenitors with fast rotation, might provide stronger GW emissions, 
and may be detectable at extra-galactic distances by the current LIGO-Virgo-KAGRA network \citep{2018ApJ...857...13P, 2020MNRAS.tmpL..27S,2020MNRAS.tmp.1245P}.  
In this paper, we present results with extreme conditions in which black hole (BH) formation occurs during the simulations,  
and we investigate the rotational effects on the dynamics of BH formation and multimessenger signals. 
BH formation in 3D CCSN simulations have been investigated by \cite{2018MNRAS.477L..80K} with a 70 $\Msun$ zero metallicity progenitor from \cite{2014ApJ...794...40T},
and an early BH formation within $t < 300$~ms postbounce is observed. 
\cite{2018ApJ...852L..19C} simulate a 40 $\Msun$ zero metallicity progenitor from \cite{2010ApJ...724..341H} and 
obtain BH formation at $\sim 0.9$~s due to fallback accretion. 
The evolution of GW frequencies from core bounce to BH formation reflect the evolution and oscillation of the central proto-neutron star (PNS), 
which are crucial probes to understand the microphysics and  supernova engine(s)
\citep{2013ApJ...779L..18C, 2018ApJ...857...13P, 2018MNRAS.477L..80K}.
However, both \cite{2018MNRAS.477L..80K} and \cite{2018ApJ...852L..19C} do not consider rotation in their simulations.
Recently, \cite{2020MNRAS.tmp.1245P} conducted simulations of three massive rotating progenitors with initial helium star masses of 18, 20, and 39 $\Msun$, 
and found that the high-frequency f/g-mode GW emissions are sensitive to rotation.  
In the present paper, we explore the rotational effect on a 40-$\Msun$ solar metallicity progenitor and investigate the impact on BH formation. 

The paper is organized as follows. 
In Section~\ref{sec_method}, we describe our simulation code, included physics, numerical schemes, 
and initial conditions. 
In Section~\ref{sec_result}, we present the results of our simulations 
and describe the rotational effects on the general evolution, shock dynamics, neutrino emissions, and GW signals.
Finally, we summarize our results and conclude in Section~\ref{sec_conclusion}.

\section{NUMERICAL METHODS} \label{sec_method}
We use the publicly available code {\tt FLASH}\footnote{http://flash.uchicago.edu} version~4 
\citep{2000ApJS..131..273F, 2008PhST..132a4046D} 
to solve the Eulerian hydrodynamics equations in multiple dimensions. 
Self-gravity is solved by the improved multipole Poisson solver of \cite{2013ApJ...778..181C} 
with a maximum multipole value $l_{\rm max} = 16$.
To mimic the general relativistic (GR) effects, we replace the monopole moment of the gravitational 
potential with an effective GR potential based on the Case A implementation that is described 
in \cite{2006A&A...445..273M} and \cite{2018ApJ...854...63O}.  
Note that \cite{2019arXiv190701138W} pointed out that the use of this effective GR potential 
might overestimate the gravitational wave frequency by $\sim 15\%$ \citep{2020PhRvL.125e1102Z}. 
We use the Isotropic Diffusion Source Approximation (IDSA, \citealt{idsa}) 
to solve for the neutrino transport of electron flavor neutrinos and 
use a leakage scheme for $\mu$ and $\tau$ flavor neutrinos \citep{2003MNRAS.342..673R}.
We use the Bruenn description for weak interactions in \cite{1985ApJS...58..771B} 
except for neutrino-electron scattering (NES). 
NES is approximated during the collapse from a parametrized deleptonization (PD) 
scheme \citep{2005ApJ...633.1042L}.  
The leakage scheme is based on the local absorption and production rates in \cite{1998ApJ...507..339H}. 
The diffusion source in our version of the IDSA solver is solved in full 3D 
and has been accelerated with GPU acceleration with {\tt OpenACC} 
\citep{2017nuco.confb0703P, 2018ApJ...857...13P, 2019JPhG...46a4001P}. 
Twenty neutrino energy bins spaced logarithmically from 3 MeV to 300 MeV are used for
electron flavor neutrinos and anti-neutrinos. 
Note that close to black hole formation, we find that the spectra of a few energy bins become noisy. 
We smooth these energy bins in order to get a smooth neutrino luminosity. 
The detailed method description and implementation of our IDSA solver is described   
in \cite{2016ApJ...817...72P, 2018ApJ...857...13P, 2018A&A...619A.118C} and \cite{2019JPhG...46a4001P}.

Our grid and hydro setup in {\tt FLASH} is nearly identical to what has been used in
\cite{2013ApJ...765...29C, 2014ApJ...785..123C, 2018ApJ...854...63O, 2016ApJ...817...72P, 2018ApJ...857...13P}.
The simulation box includes the inner 10,000~km of a progenitor in 3D Cartesian coordinates and 
we use 9 levels of adaptive mesh refinement (AMR) in our simulations, giving a cell width of $0.488$~km in the finest zone.
To save computing time, we reduce the AMR level as a function of the distance to the center, 
giving an effective angular resolution of $\sim 1^\circ$. 
A power law profile in radius is used as outer boundary condition for density and velocity to mimic 
the stellar envelope \citep{2013ApJ...765...29C}. 
We use the $v$-constant rotation formula of \cite{1985A&A...146..260E, 1985A&A...147..161E} 
to model the rotation of a progenitor,
\begin{equation}
\label{eq_rot}
\Omega(r) = \frac{\Omega_0}{1+(r/A_0)^2},
\end{equation} 
where $\Omega$ is the angular velocity, $r$ is the cylindrical radius, $A_0 = 1,000$~km is a scaling constant we fixed in this study, 
and $\Omega_0$ is a free parameter to model different rotational speeds. 
Note that this rotation profile uses the cylindrical radius instead of the spherical radius. 
See \cite{1997A&A...321..465M} and \cite{2000ApJ...528..368H} for discussions of ``shellular'' vs ``cylindrical rotation.''

We use the 40-\Msun progenitor at zero-age main sequence with solar metallicity (s40) from \cite{2007PhR...442..269W} as initial condition. 
We also use the Lattimer \& Swesty equation of state (EoS) with incompressibility $K=220$~MeV 
(LS220, \citealt{1991NuPhA.535..331L}). 
It should be noted that although LS220 EoS is not completely ruled out, 
recent studies suggest that the LS220 EoS does not fulfill the constraints from
chiral effective field theory \citep{2013PhRvC..88b5802K} and it uses the single nucleus approximation for heavy nuclei. 
\citet{2011ApJ...730...70O} have shown that the s40 progenitor with the LS220 EoS has the shortest 
BH formation time among valid EoSs and the progenitors in the 2007 progenitor set in 
\citet{2007PhR...442..269W}. 
This is further confirmed by 2D simulations with different EoSs \citep{2018ApJ...857...13P}.
Thus, we use this combination to save computing time in 3D simulations.
Note that we ignore magnetic fields in this paper for simplicity, although it is considered as an import ingredient to explain some long gamma-ray bursts, 
 especially for cases with fast rotating progenitors 
 \citep{2012ApJ...750L..22W, 2014ApJ...785L..29M, 2015Natur.528..376M, 2020MNRAS.492.4613O, 2020arXiv200302004K}.
 
We conduct three 3D simulations from the onset of core collapse with $\Omega_0=0, 0.5$ and $1.0$ rad~s$^{-1}$. 
The PD scheme is used during collapse to update the electron fraction, entropy, 
and the momentum transfer from neutrino stress \citep{2005ApJ...633.1042L}. 
In addition, we include the non-rotating 2D simulation with LS220 EoS from \cite{2018ApJ...857...13P} as a comparison.
In this paper, we denote these three 3D models from non-rotating to $\Omega_0=1$~rad~s$^{-1}$ 
as NR (non-rotating), SR (slowly-rotating), and FR (fast-rotating), and the non-rotating 2D model as NR-2D. 

\section{RESULTS} \label{sec_result}

\begin{figure*}
	\epsscale{1.2}
	\plotone{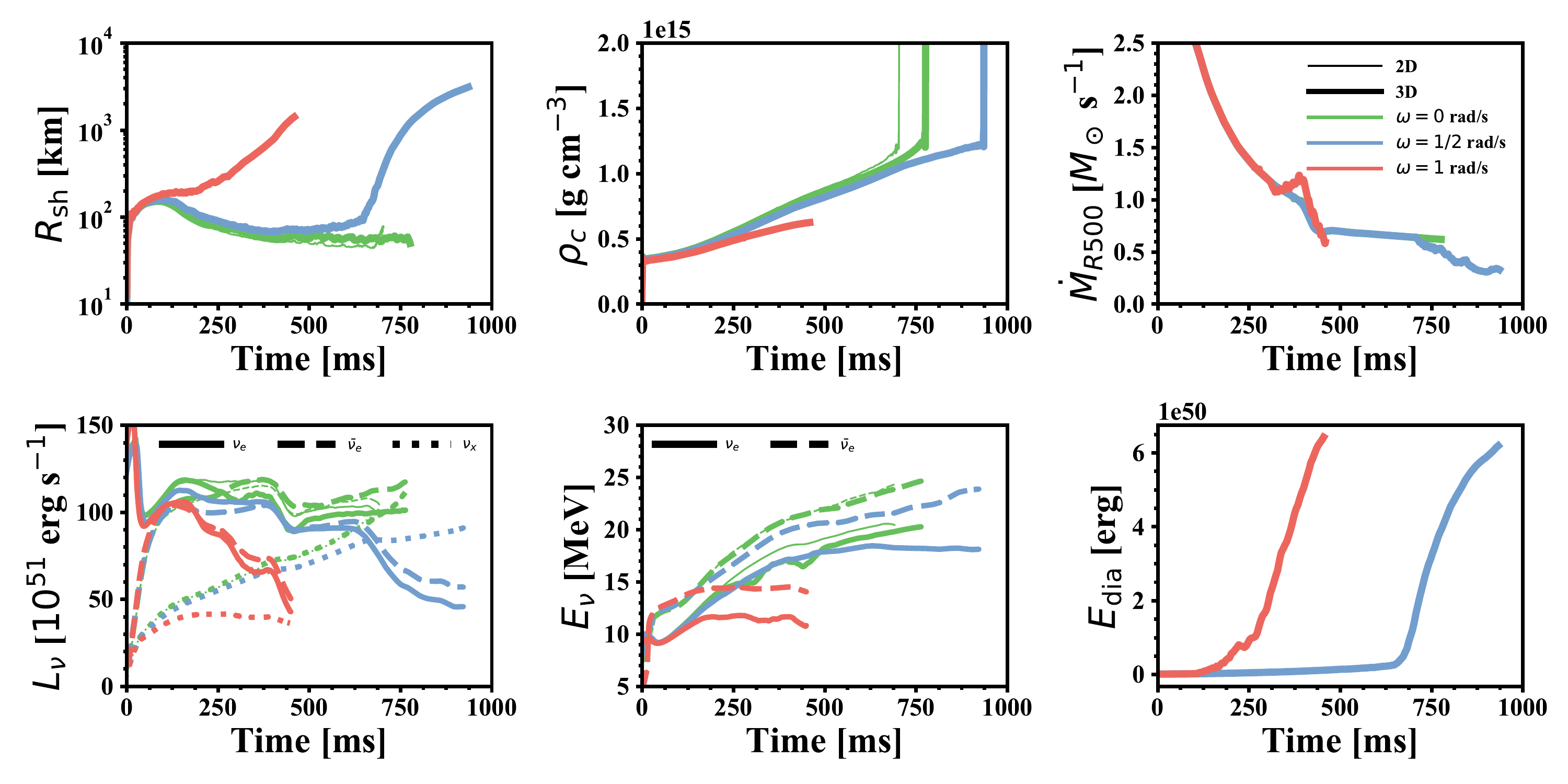}
	\caption{\label{fig_time}
	Time evolution of averaged shock radius, central density, mass accretion rate (measured at $r=500$~km), 
	neutrino luminosities, neutrino mean energies, and diagnostic explosion energy. 
	Different colors represent simulations with different rotational speeds. 
	The thin green line shows a 2D counterpart of the non-rotating model in \cite{2018ApJ...857...13P}. 
	A moving averaged filter with a window of 3~ms is applied to the neutrino luminosity 
	and mean energy to reduce noise.}
\end{figure*}

In this section, we present the results of our 3D simulations with different initial rotational speeds.
We first describe the general properties of our simulations and then focus on 
discussions of the standing accretion shock instability (SASI), angular momentum re-distribution,  
and multimessenger signals. 

%
%
%

\subsection{General properties and black hole formation}

The NR and NR-2D models give an almost identical bounce time 
because the stellar core remains nearly spherically symmetric during core collapse.
The core bounce in the SR and FR models is delayed relative to the NR cases by $\sim4$~ms and $\sim13$~ms, respectively, due to the effects of rotation. 
Figure~\ref{fig_time} shows the time evolution of averaged shock radius, central density,
mass accretion rate (measured at 500~km), neutrino luminosity, mean energy, 
and diagnostic explosion energy of all our models. 
All models show a similar central density at bounce, but the later evolution depends strongly on the rotational rate:
the faster the rotational speed, the slower the rate of increase in the central density. 
The NR model (thick green lines) is a failed SN, and a BH is formed at 776~ms postbounce without shock revival. 
Since our code uses an approximated GR treatment, we could not follow the simulation up to the appearance of an event horizon.
In this paper, we define BH formation when the central density of a PNS suddenly and rapidly increases, reaching the upper density limit of
the nuclear EoS table ($\rho_{\rm max} \sim 3 \times 10^{15}$~g~cm$^{-3}$). 
This can be seen in the final part of the central density evolutions of green and blue lines in Figure~\ref{fig_time}. 
We terminate a simulation when this criterion is achieved.  
The baryonic mass of the PNS at this time is about $2.6 M_\odot$.
Relative to the 2D counterpart (model NR-2D) in \cite{2018ApJ...857...13P}, the BH formation time in the 3D NR model is delayed by $\sim 72$~ms, but the neutrino luminosities and mean energies are still very similar.

\begin{figure*}
	\epsscale{1.0}
	\plotone{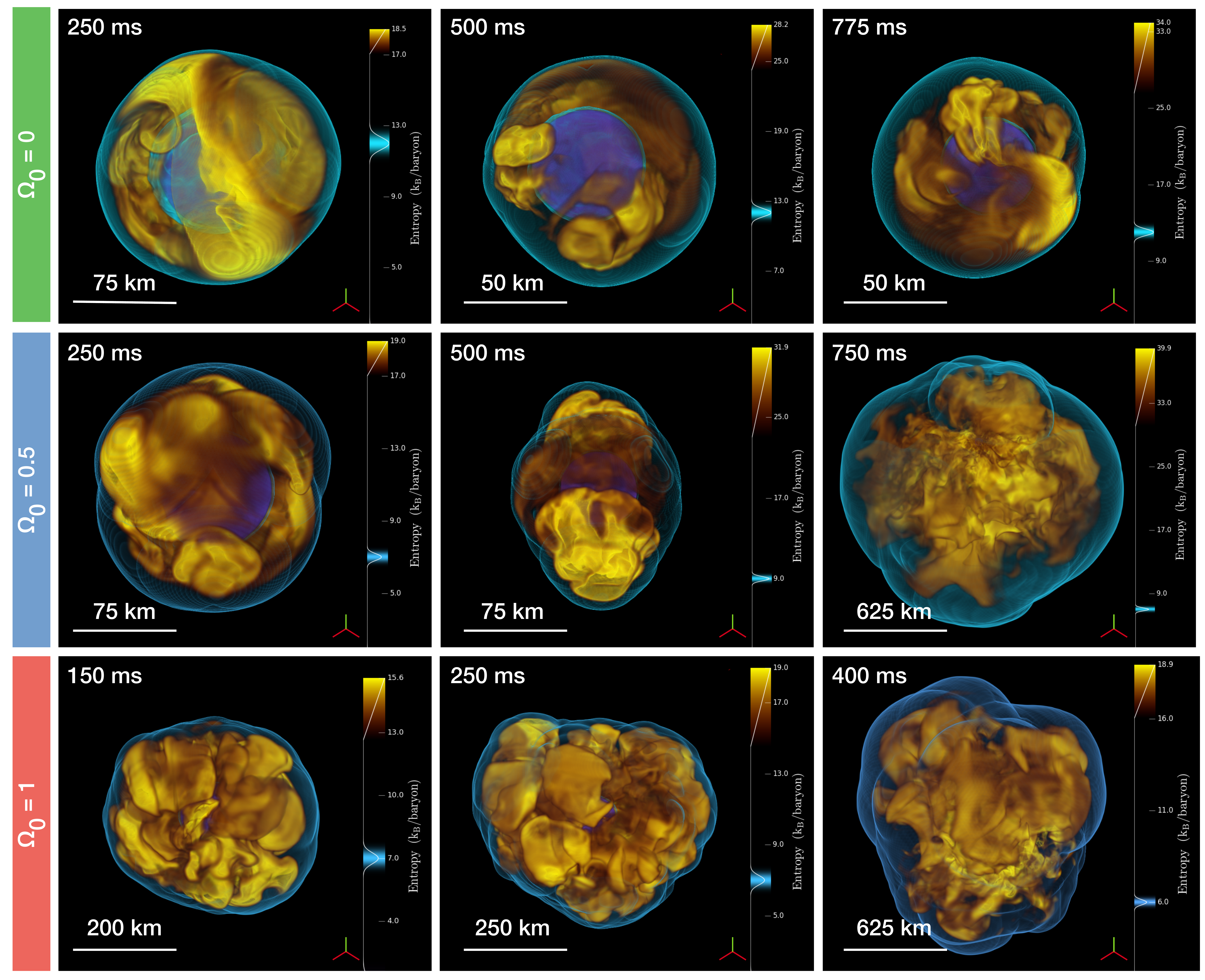}
	\caption{\label{fig_vr}
	Volume rendering of the entropy, density, and shock front at different postbounce time (column) 
	and with different initial rotational speed (row). 
	The yellow color represents the entropy, the purple color shows the surface of PNS, 
	and the thin blue layer indicates the shock front.
	Positive $z$-axis (the green ticker) is pointing to the up direction,  
	but the angular momentum vector is pointing to the down direction.}
\end{figure*}

At $\sim 50$~ms before BH formation, the NR-2D model shows signs of explosion as the shock starts to expand rapidly.
We do not see this feature in model NR. This difference can be understood by the presence of 
the third dimension in model NR, and the fact that the NR-2D model has a higher resolution in the gain region 
than the corresponding 3D model (NR), resulting a higher neutrino luminosity in the NR-2D model.
Figure~\ref{fig_vr} shows volume rendering plots of entropy at different time and with different initial rotational speeds.

The SR model behaves similarly to the NR model in the first 400~ms postbounce, 
but subsequently shock revival is achieved in the SR model at $t>650$~ms postbounce. 
Once the shock has revived, the diagnostic explosion energy quickly increases to $> 6\times 10^{50}$~erg 
at the end of our simulation when a BH is formed. 
BH formation takes place at $936$~ms postbounce. 
This shock revival is similar to the full~GR simulation of a 70~$\Msun$ (Z70) progenitor in \cite{2018MNRAS.477L..80K},
but in our case, the explosion is not only aided by the strong convection behind the shock but also by rotation. 

\begin{figure*}
	\epsscale{0.45}
	\plotone{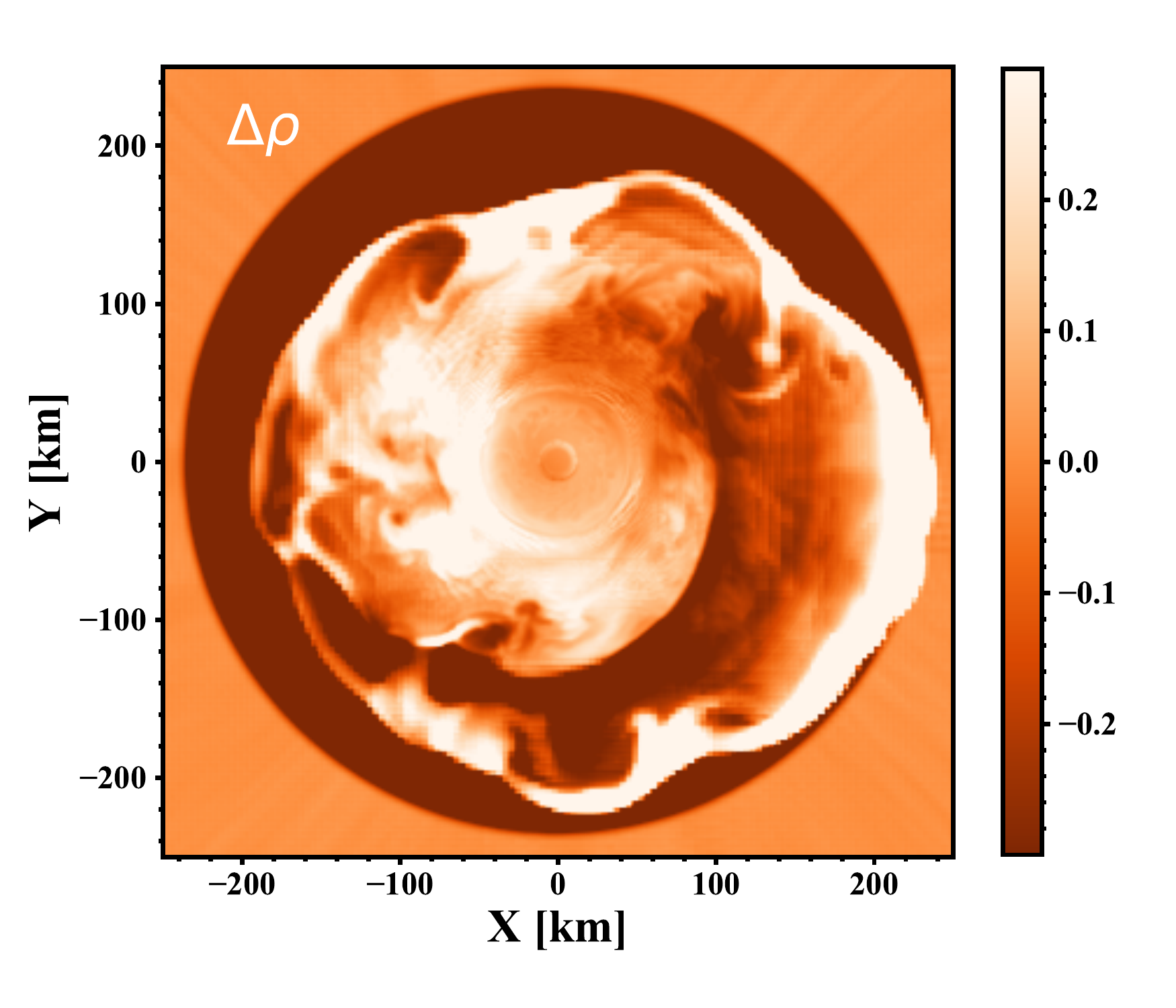}
	\plotone{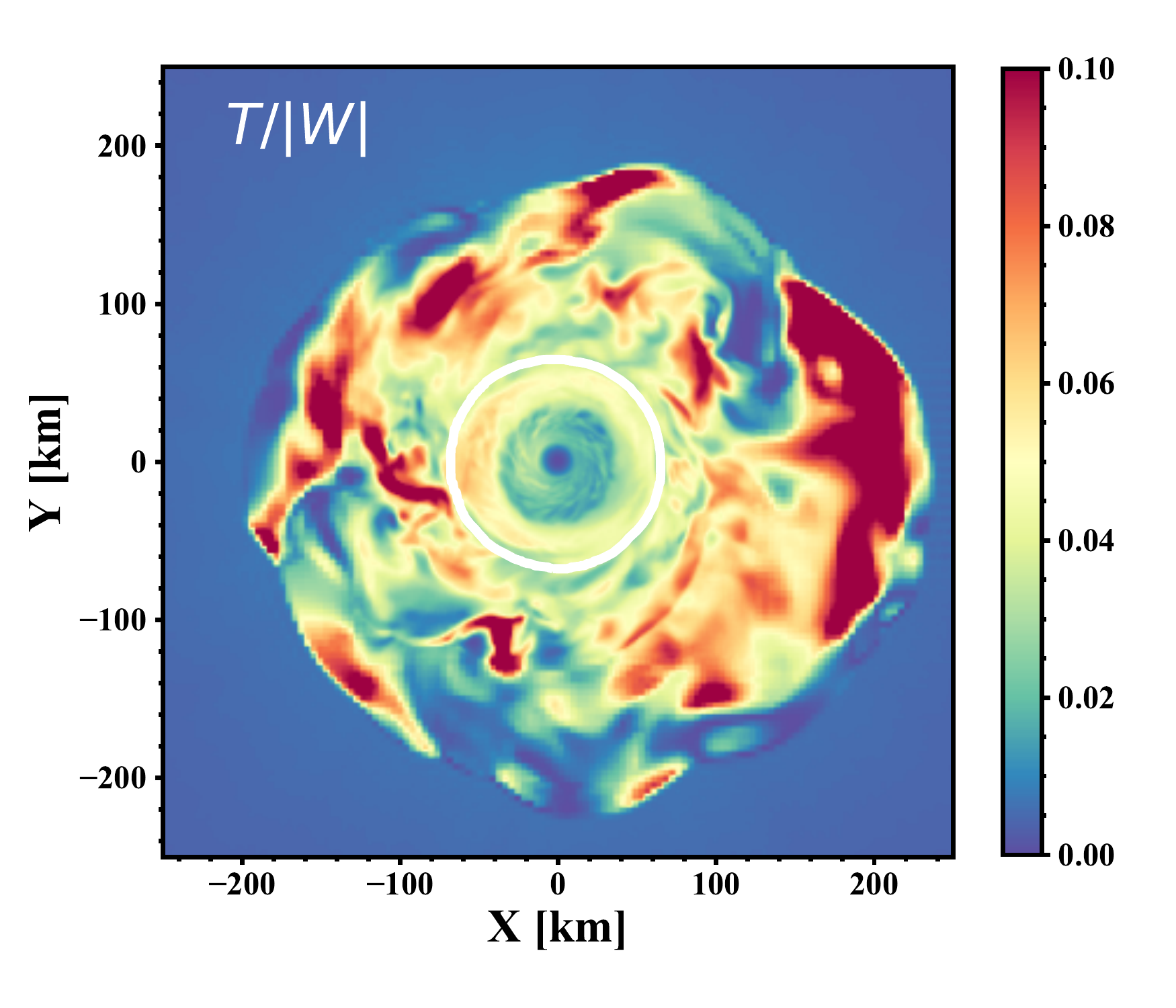}
	\caption{\label{fig_arm}
	Left: The density variation (defined as in Equation~\ref{eq:spiral}) of the FR model at 120~ms postbounce.
	Right: the ratio of rotational to gravitational energy $T/|W|$ on the rotational plane of FR model at 120~ms postbounce. }
\end{figure*}

Unlike the NR and SR models, the FR model experiences a fast shock expansion and 
explosion very early at $\sim 250$~ms postbounce. The diagnostic explosion energy reaches $\sim 6 \times 10^{50}$~erg at 
$\sim 460$~ms postbounce when the averaged shock front is above 1,000~km. 
We terminate the FR simulation at $\sim460$~ms postbounce due to the inhibitively large computational cost in the shocked region at large radii.
We do not find BH formation within $\sim 460$~ms postbounce.
The FR model also has the lowest neutrino mean energies and luminosity due to the early, fast shock expansion. 
In Figure~\ref{fig_arm}, we show the density variation as defined in \cite{2005ApJ...625L.119O, 2010A&A...514A..51S, 2016MNRAS.461L.112T} at 120~ms postbounce, 
\begin{equation}
\label{eq:spiral}
\Delta \rho \equiv \frac{\rho - \left< \rho \right>}{ \left< \rho \right>},
\end{equation}
where angle brackets stand for the averaging over angles. 
A one-arm $(m=1)$ spiral instability has developed, that helps to transfer the angular momentum outward and eventually leads to an early explosion. 
The corresponding $T/ |W|$ ratios are plotted in the right panel in Figure~\ref{fig_arm}.
The values of $T/ |W|$ around the PNS and the gain region are between $4\%$ and $10\%$, 
which are typical values for the so-called low $T/ |W|$ instability that has been described in 
\cite{2005ApJ...625L.119O, 2007PhRvL..98z1101O, 2014PhRvD..89d4011K, 2016MNRAS.461L.112T} and \cite{2020MNRAS.tmpL..27S}. 
\cite{2018MNRAS.475L..91T} found a quasi-periodic modulation of anisotropic neutrino signals, 
which is associated with the one-arm spiral flow, in a rapid-rotating supernova simulation. 
We are unable to diagnose this modulation of the neutrino signals in our FR model since our free-streaming neutrinos are angle averaged.   

%
%

\subsection{PNS Rotation and SASI}

\begin{figure*}
	\epsscale{1.0}
	\plotone{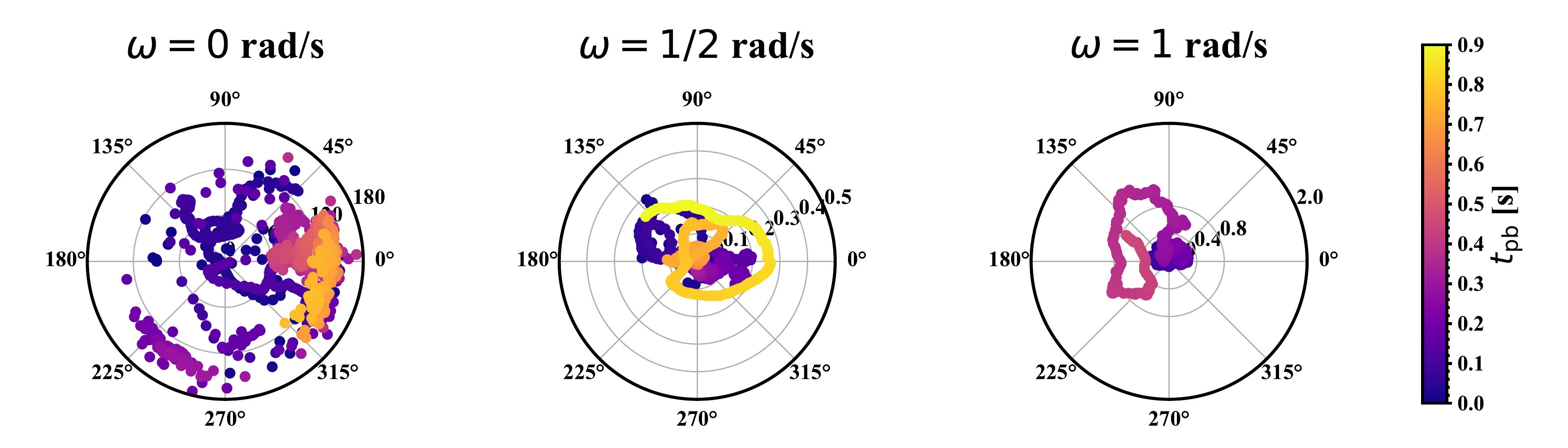}
    \caption{\label{fig_pns_axis}
	Time evolution of the direction of the rotational axes of the PNS in models with different initial rotational speeds.
	The origin of this polar plot is pointed to the direction of the negative z-axis in the simulation box.
	The distance to the origin represents the azimuthal angle ($\phi$) of the rotational axis, 
	and the phase angle shows the polar angle ($\theta$) of the rotational axis.
	Colors from blue to yellow represent the simulation time after core bounce. 
	Note that the azimuthal angle limits from left to right panels are $180^\circ$, $0.5^\circ$, and $2.0^\circ$.}
\end{figure*}

\begin{figure*}
	\epsscale{1.0}
	\plotone{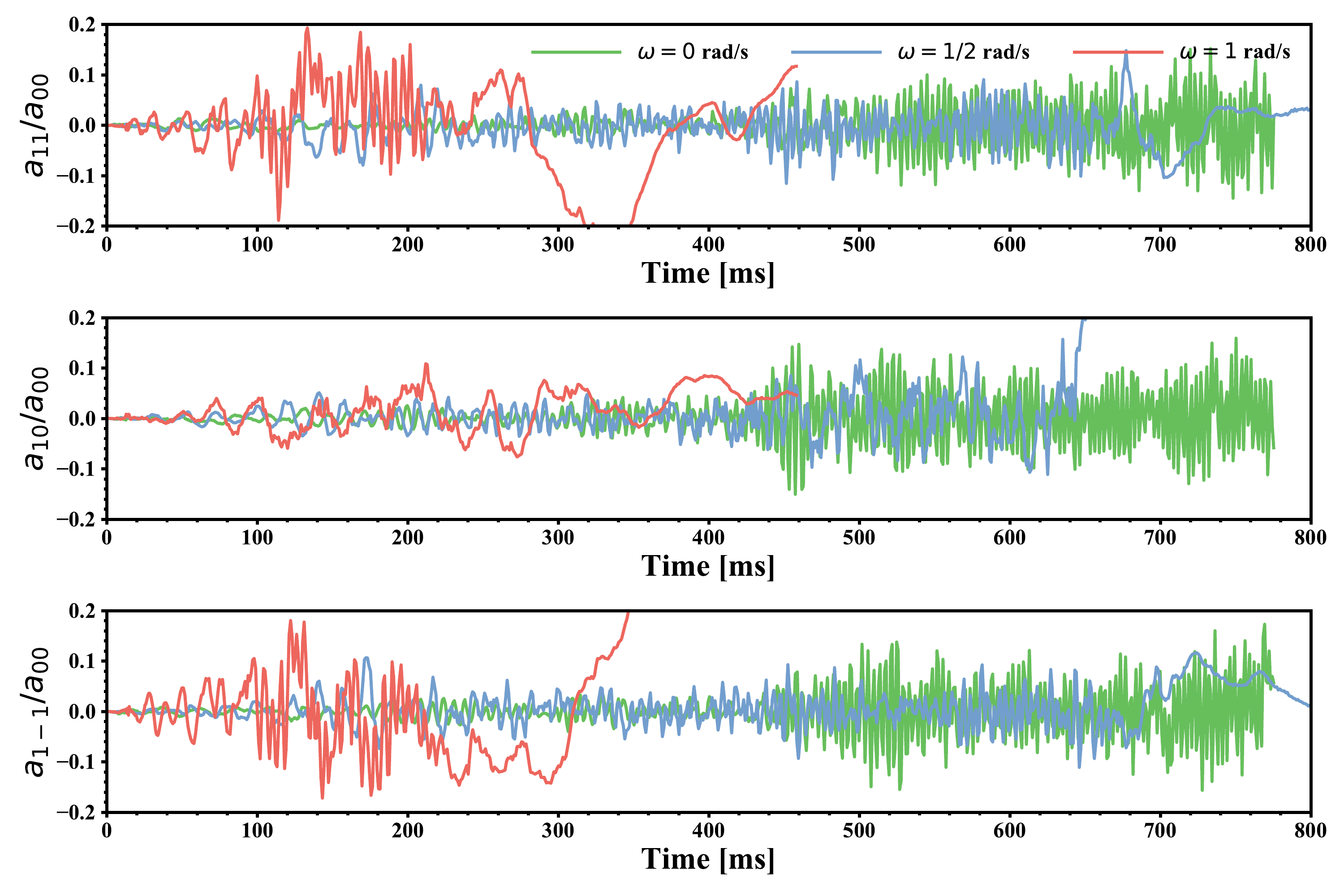}
	\caption{\label{fig_sasi}
	Time evolution of SASI amplitudes in different directions. 
	Different colors represent different initial rotational speed. 
	Amplitudes are normalized by the averaged shock radius ($a_{00}$).}
\end{figure*}

\begin{figure*}
	\epsscale{0.7}
	\plotone{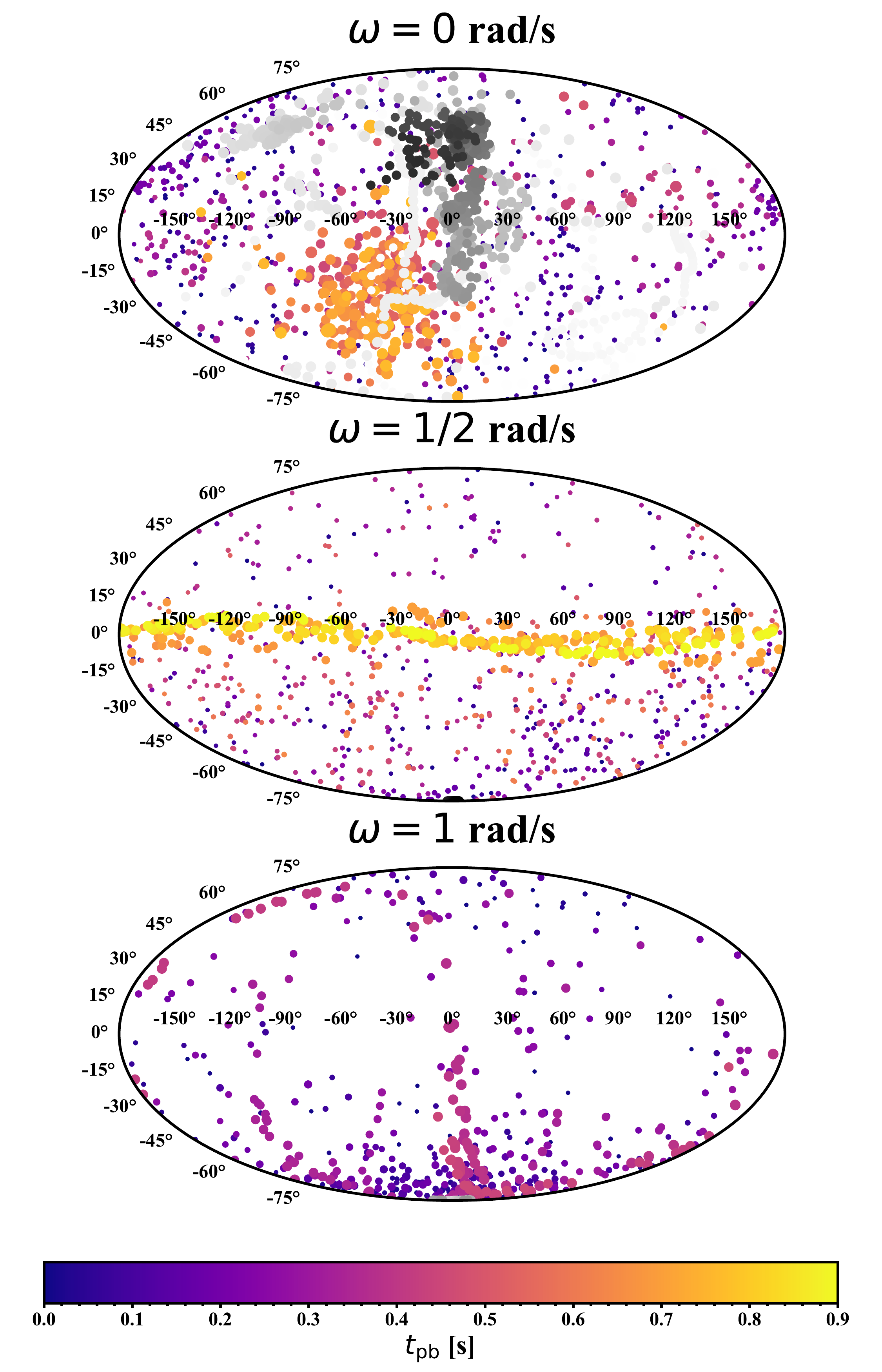}
	\caption{\label{fig_spns}
    Distribution of PNS rotational axes (gray/black dots) and SASI axes (colored dots). 
    Colors from blue (gray) to yellow (black) represent the simulation time after core bounce.
    The dot size represents the normalized strength.
    Note that in the middle (model SR) and bottom (model FR) panels, 
    the rotational axes point nearly to the south pole and therefore show only one dot at the lower edge of the plot.} 
\end{figure*}

Figure~\ref{fig_pns_axis} shows the time evolution of the direction of the angular momentum of the PNS with different initial rotational speeds.
The magnitude of angular momentum in the NR model (the left panel) is nearly zero during early postbounce,
and therefore the direction of the angular momentum vector is pointing in a random direction 
(the blue and purple dots in the left panel in Figure~\ref{fig_pns_axis}). 
However, following bounce, a preferred direction is excited due to convection and SASI activity.
Spiral modes of SASI could help to redistribute angular momentum and  
transport angular momentum to the PNS,
resulting in a spin-up of the PNS \citep{2007ApJ...656..366B, 2007Natur.445...58B, 2018ApJ...865...81O}. 
The SR (FR) model shows a similar effect from the spiral SASI resulting in a small precession of the angular momentum vector of $\sim 0.3^\circ$ ($\sim 1^\circ$).

To further investigate the angular momentum transport between SASI and PNS, 
we follow \cite{2006ApJ...642..401B, 2014ApJ...785..123C, 2017MNRAS.468.2032A} 
to evaluate SASI directions by decomposing the shock front into spherical harmonics.
Figure~\ref{fig_sasi} shows the normalized SASI amplitudes, $a_{lm}$, in different axes, corresponding to $l=1$, $m=1,-1, 0$ modes of spherical harmonics, 
where $a_{lm}$ is defined by 
\begin{equation}
a_{lm} = \int_{\it S} R_s(\theta, \phi) Y_{lm}(\theta, \phi) d \Omega,
\end{equation}
with $R_s$ being the shock location, $\theta$ and $\phi$ the two angles in spherical coordinates, $\Omega$ the solid angle, and 
$Y_{lm}$ the Laplace's spherical harmonics.

The SASI direction vector is then defined by the vector product of two SASI vectors separated by a short time interval $\Delta t \sim 1$~ms, 
$ \vec{\boldsymbol{A}}_{SASI} = \left[a_x(t) \hat{x} + a_y(t) \hat{y} + a_z(t) \hat{z} \right] \times \left[a_x(t+\Delta t) \hat{x} + a_y(t+\Delta t) \hat{y} +(a_z(t+\Delta t) \hat{z}\right]$,
where $\hat{x}, \hat{y}$, and $\hat{z}$ are unit vectors in Cartesian coordinates. 
Figure~\ref{fig_spns} compares the SASI directions with the angular momentum vectors of PNSs. 
In the NR model, the SASI vector and angular momentum vector are pointing to a random direction early on, 
but converge to narrow and opposite directions after $\sim 500$~ms postbounce, 
suggesting that the PNS received its angular momentum from the SASI (See also \citet{2007Natur.445...58B, 2018ApJ...865...81O}).  
A similar behavior is not apparent in model SR and FR since the initial angular momentum is much higher than the contribution from SASI.  
In addition, the SASI directions in the SR and FR models have a precession due to rotation. 
The SR model has a precession angle of nearly $90^\circ$ can be understood 
from its initial shock expansion toward the pole direction at $\sim 500$ ms postbounce (see Figure~\ref{fig_vr}). 
However, the FR model has a precession angle irrelevant to its PNS spin.

\begin{figure*}
	\epsscale{1.0}
	\plotone{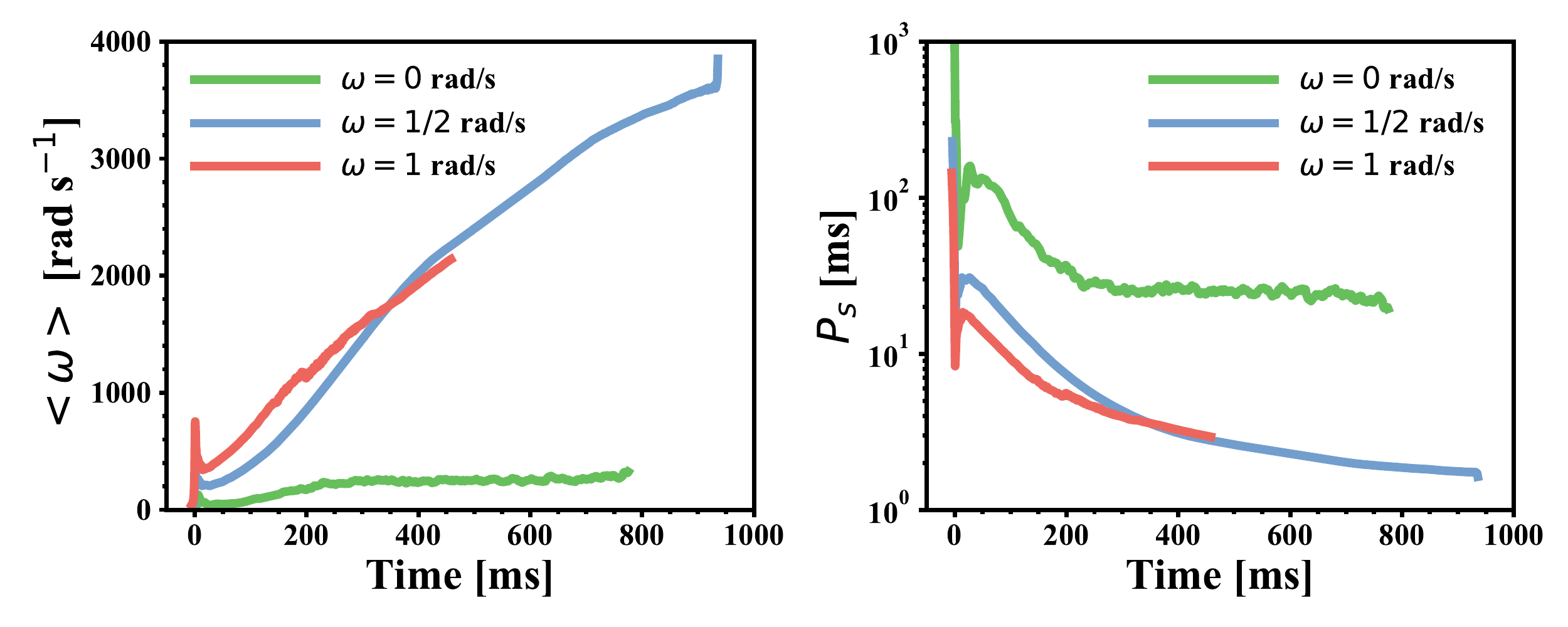}
	\caption{\label{fig_spin}
	Evolution of PNS angular velocity and rotational periods. Different colors represent different initial rotational speed. }
\end{figure*}

The contraction of the PNS will further spin it up.
Figure~\ref{fig_spin} shows the evolution of PNS rotational speed and rotational periods.
Our results suggest that even a non-rotating progenitor could end in a rotating PNS after a supernova explosion.
The SR and FR models naturally form a rotating PNS with a rotational period $\sim 30$ms, 
which can result in a millisecond pulsar if the PNS were endowed with a magnetic field. 
Note that we analyze the conservation of angular momentum by comparing the difference of angular momentum at different time within an enclosed mass of $2.5 \Msun$.
The non-conservation of angular momentum due to numerical dissipation is $\lesssim 3\%$ during the collapse 
but could be as high as $\lesssim 15\%$ near to BH formation.

If we assume that the specific angular momentum is conserved during black hole formation, 
we can estimate black hole spin parameters of $a=J/M = 0.046$ and $0.52$ for the NR and SR models, respectively. 
Note that the NR model is a failed SN, and therefore the subsequent fallback will cancel the BH's angular momentum resulting in a non-rotating BH.
However, if a similar non-rotating progenitor does explode, the spinning remnant BH might have a similar spin parameter as our NR model.
We could also estimate the remnant BH mass $M_{\rm BH} = 3.46 \Msun$ of the SR model by evaluating the bound mass at the end of our simulation, 
assuming that all the material outside the simulation box will be ejected by the explosion.
Since the FR does not undergo PNS collapse to a BH within our simulation time, we can only provide a lower limit of 
$a=0.93$ for an eventual BH resulting from the FR model.

%
%
\subsection{GW Emission}


 
\begin{figure}
	\epsscale{1.0}
	\plotone{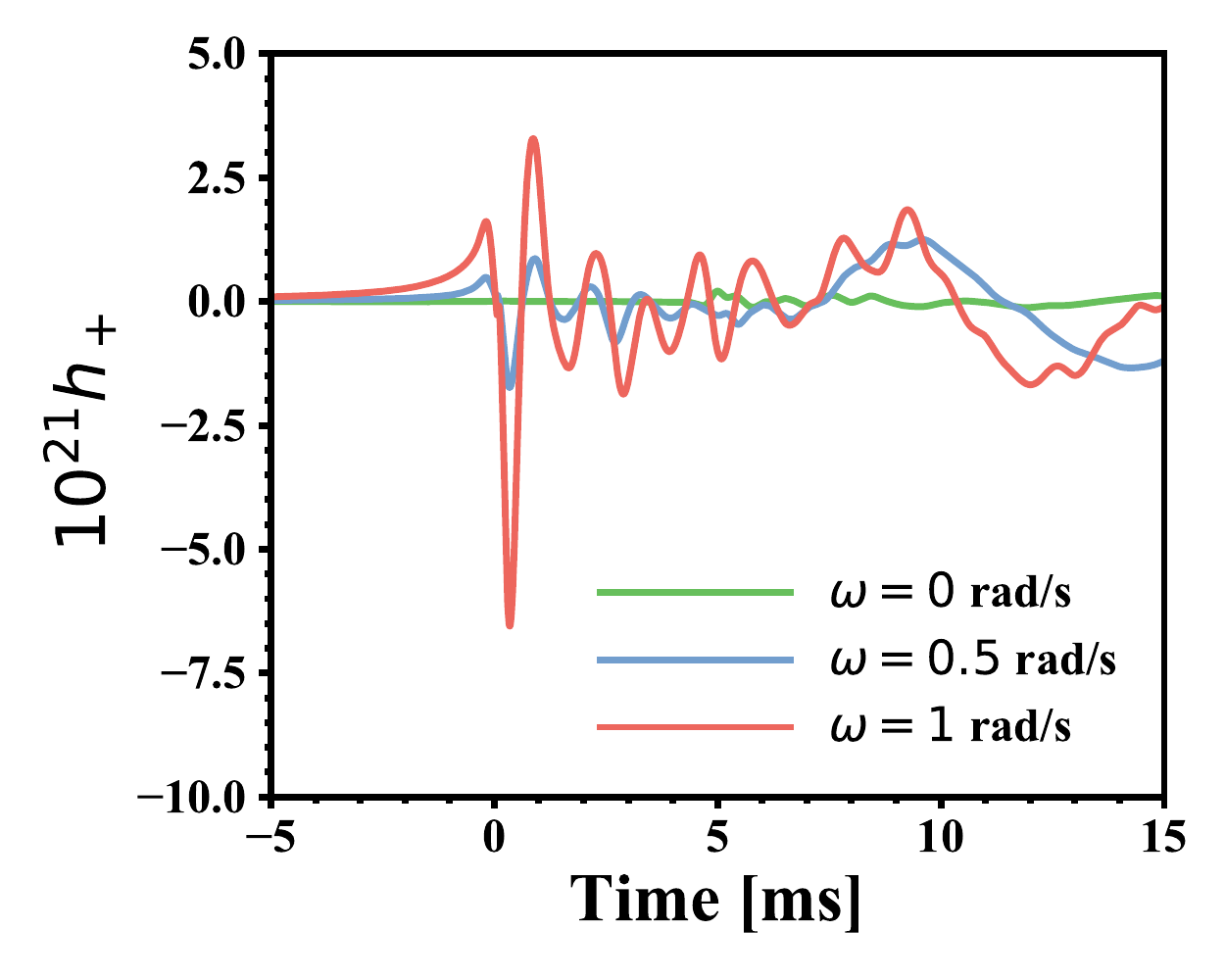}
	\caption{\label{fig_bounce}
	Gravitational wave signals at around core bounce. Different colors represent different initial rotational speed.}
\end{figure}

We follow the formulation in \cite{2008A&A...490..231S, 2009ApJ...707.1173M} and \cite{2018ApJ...857...13P} to 
evaluate the gravitational wave strains based on the the first time derivative of the mass quadrupole moment and assume a fixed distance of 10~kpc to an observer. 
The second time derivative of the mass quadrupole moment is derived by a finite difference in post-processing.
Figure~\ref{fig_bounce} shows the ``plus'' polarization gravitational waveforms of the three models at around core bounce, 
assuming an observer on the equator. 
The bounce signals are strongly correlated with the rotational speed and the amount of angular momentum in the core, 
and are consistent with what has been described in \cite{2008PhRvD..78f4056D, 2017PhRvD..95f3019R} and \cite{2019ApJ...878...13P}.

\begin{figure*}
	\epsscale{0.5}
	\plotone{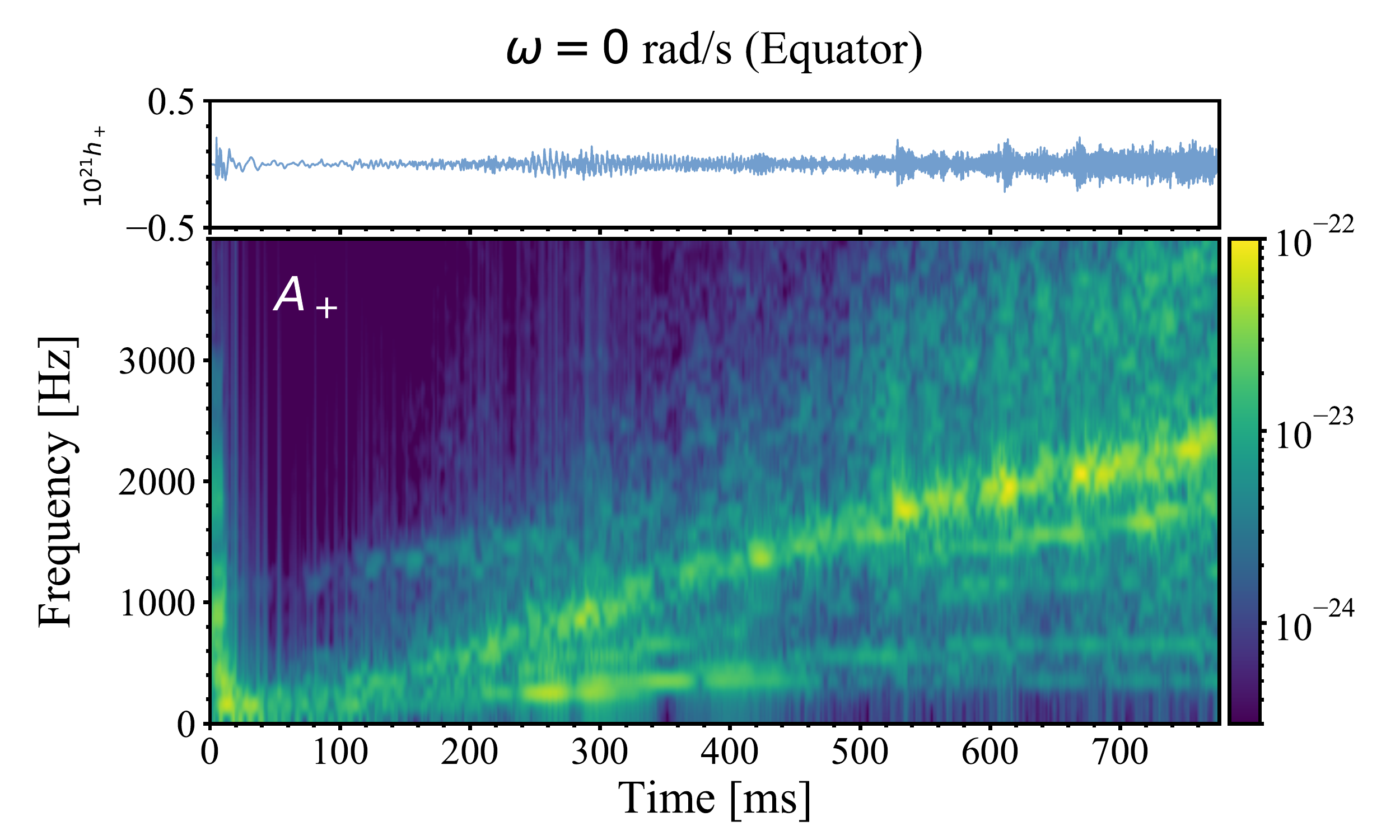}
	\plotone{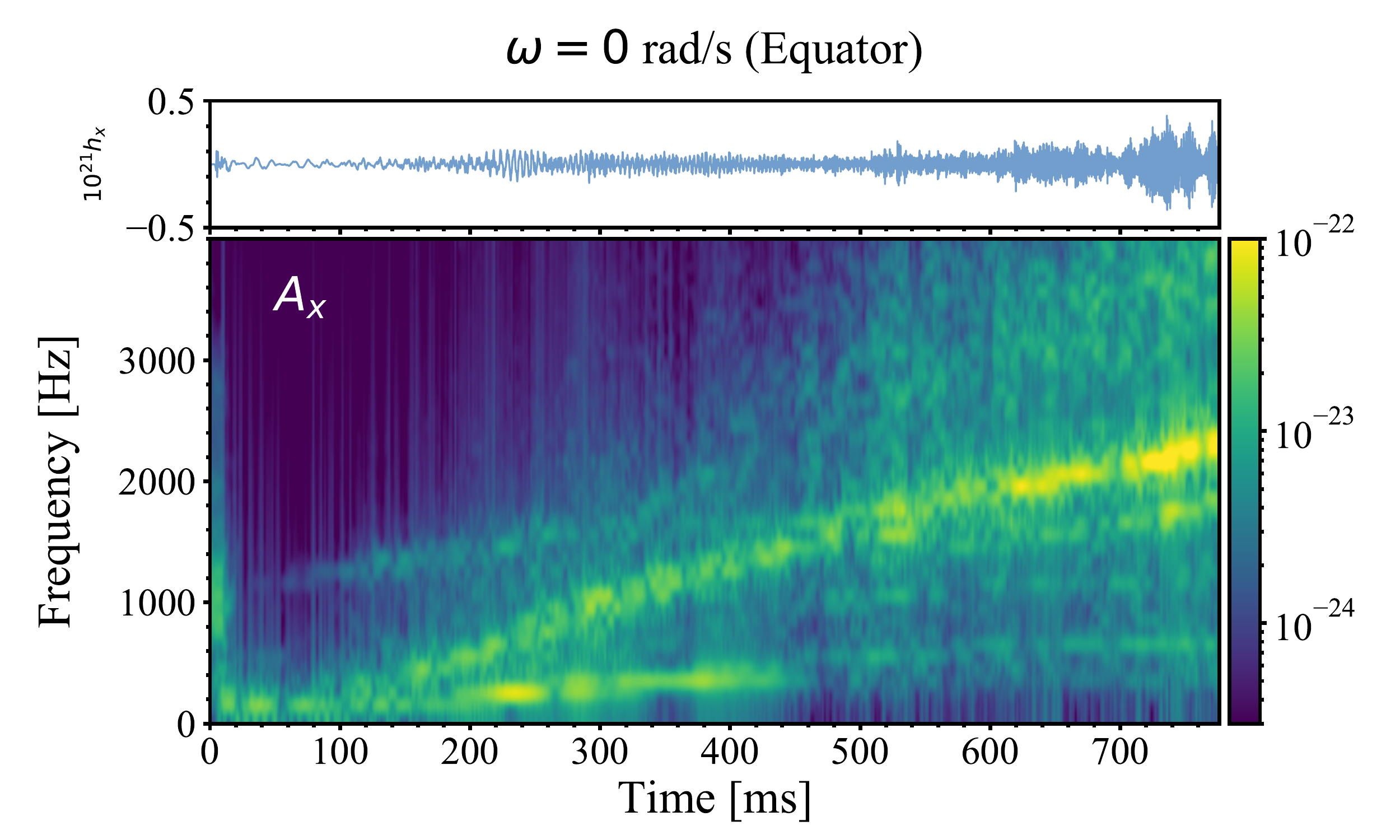}

	\plotone{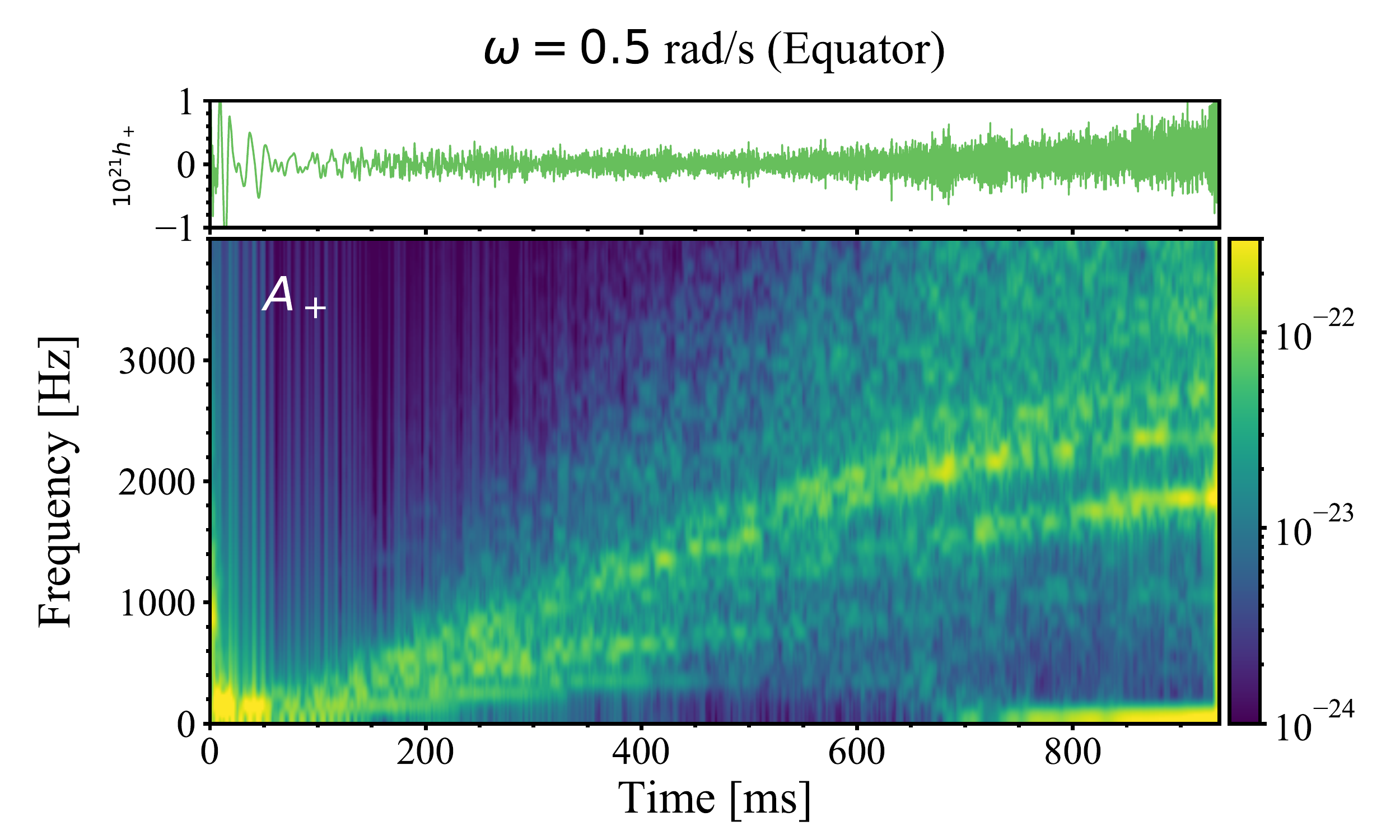}
	\plotone{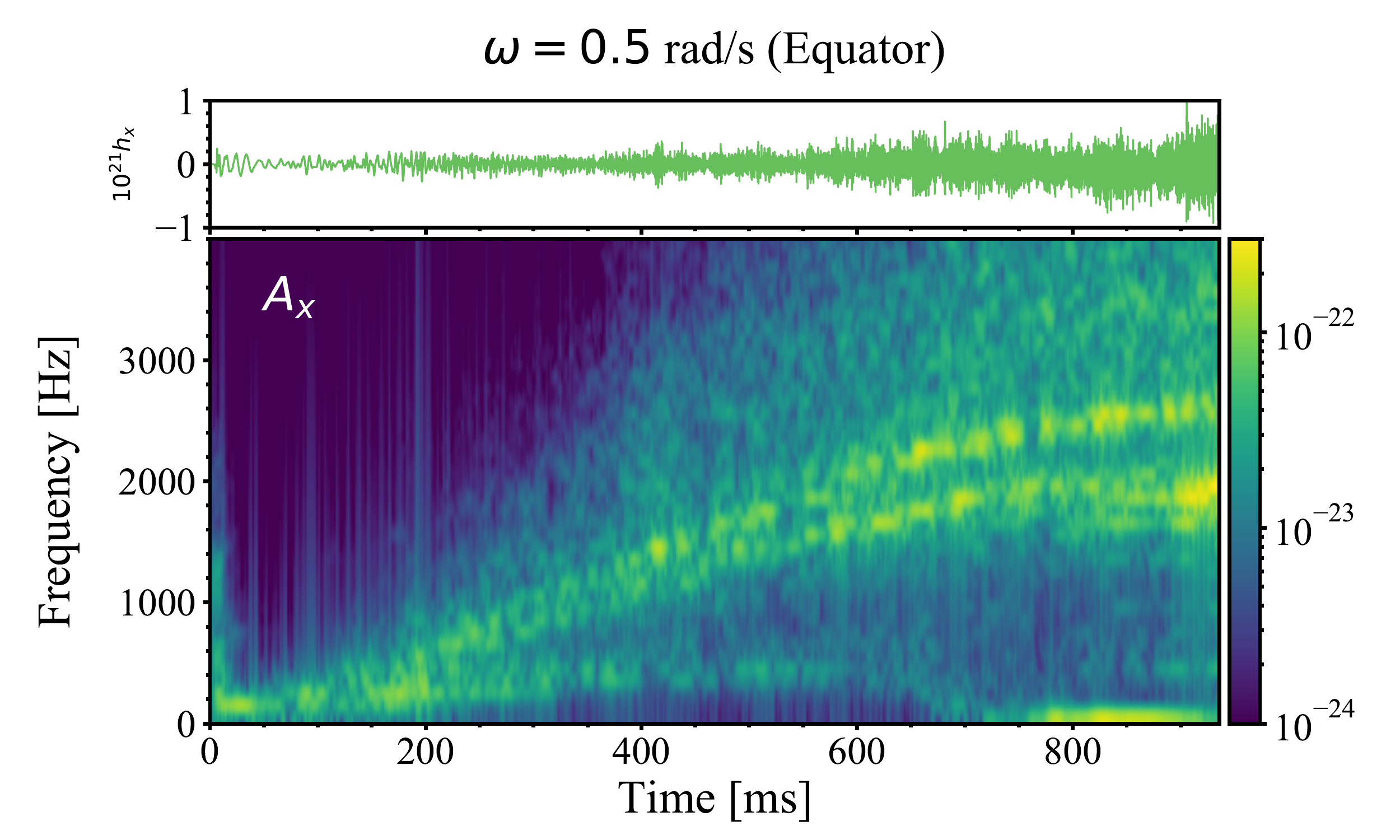}

	\plotone{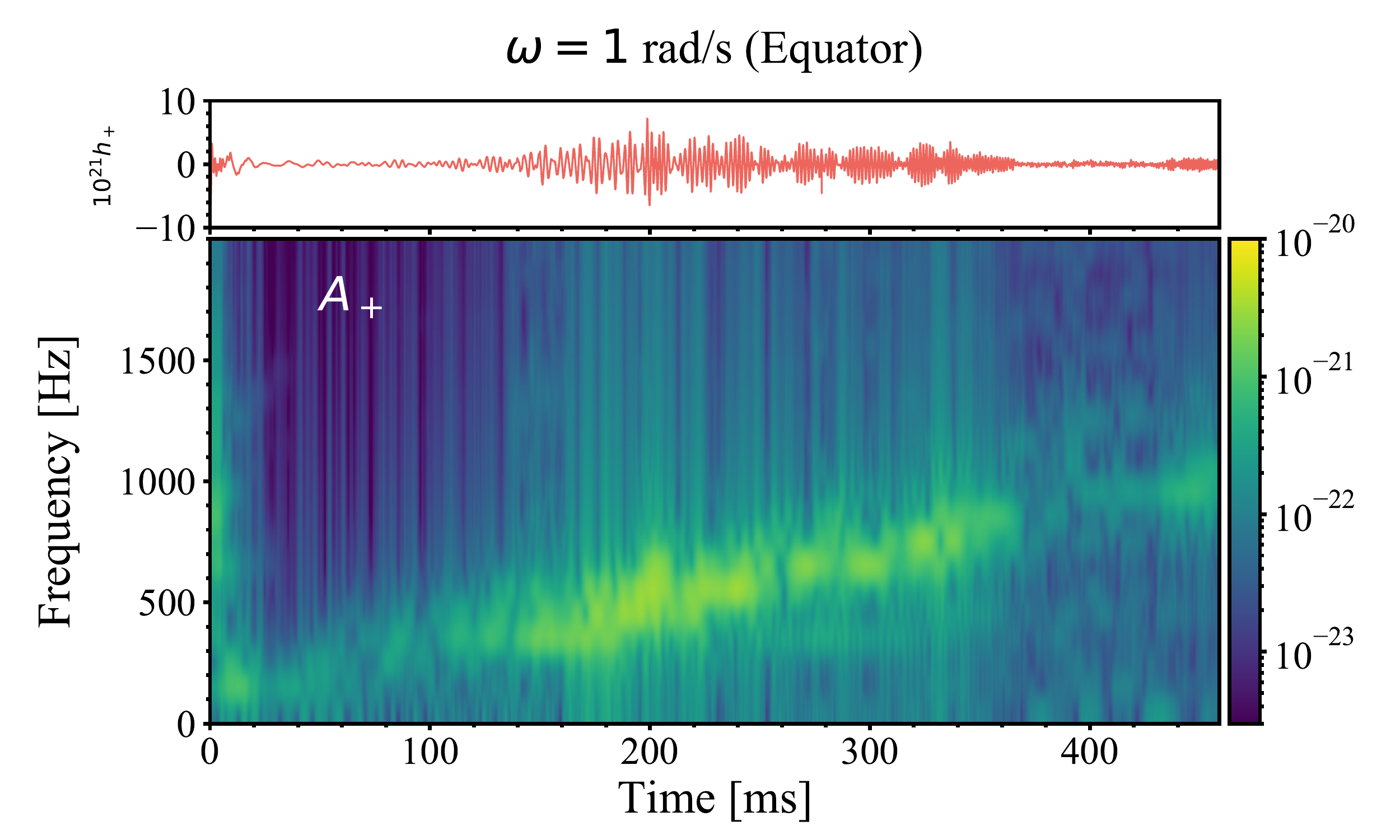}
	\plotone{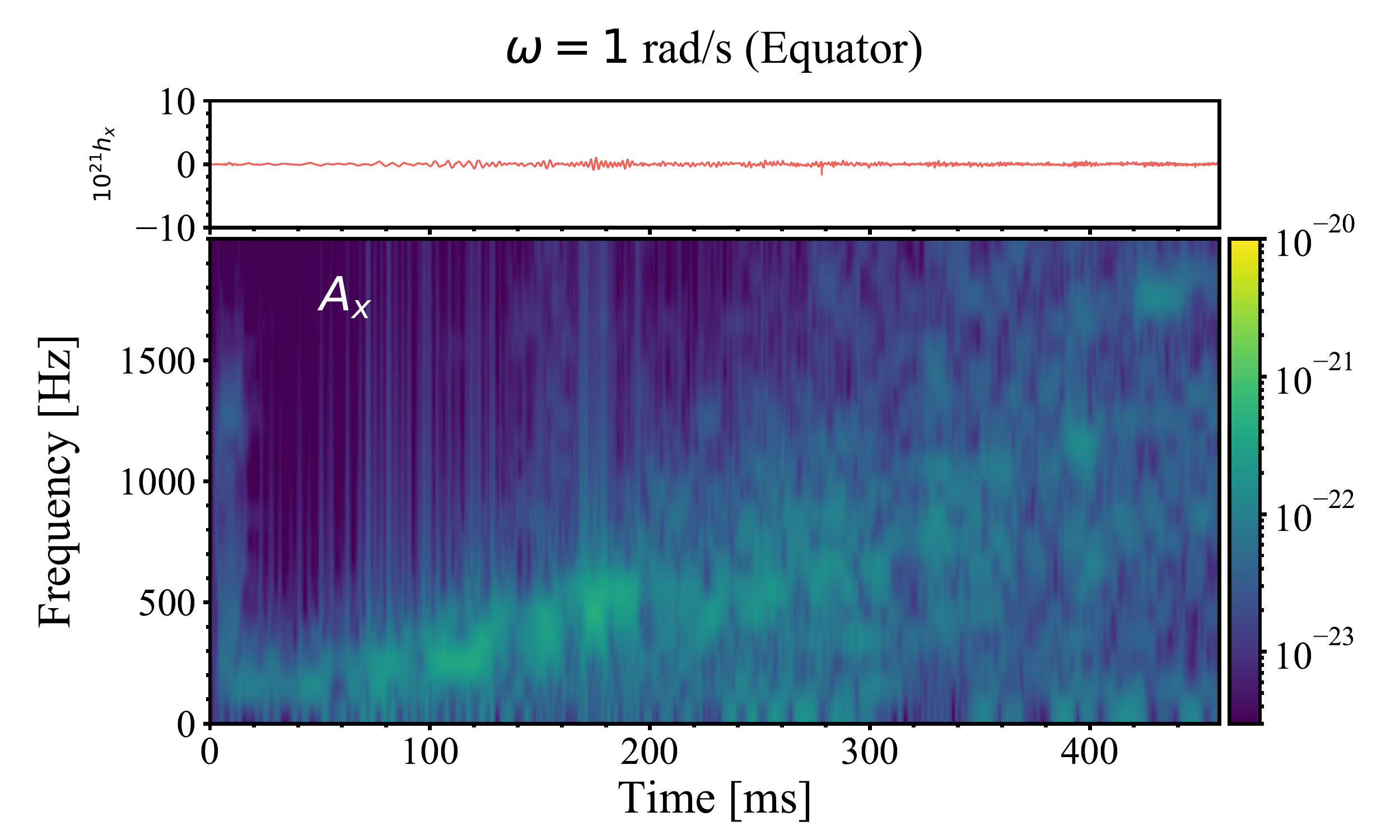}
	
	\caption{\label{fig_gw_equator}
	Spectrograms of GW strains after core bounce, as observed from the equatorial direction. 
	Different rows represent models with different initial rotational speed and different columns indicate signals with different polarization. }
	
\end{figure*}

\begin{figure*}
	\epsscale{0.5}
	\plotone{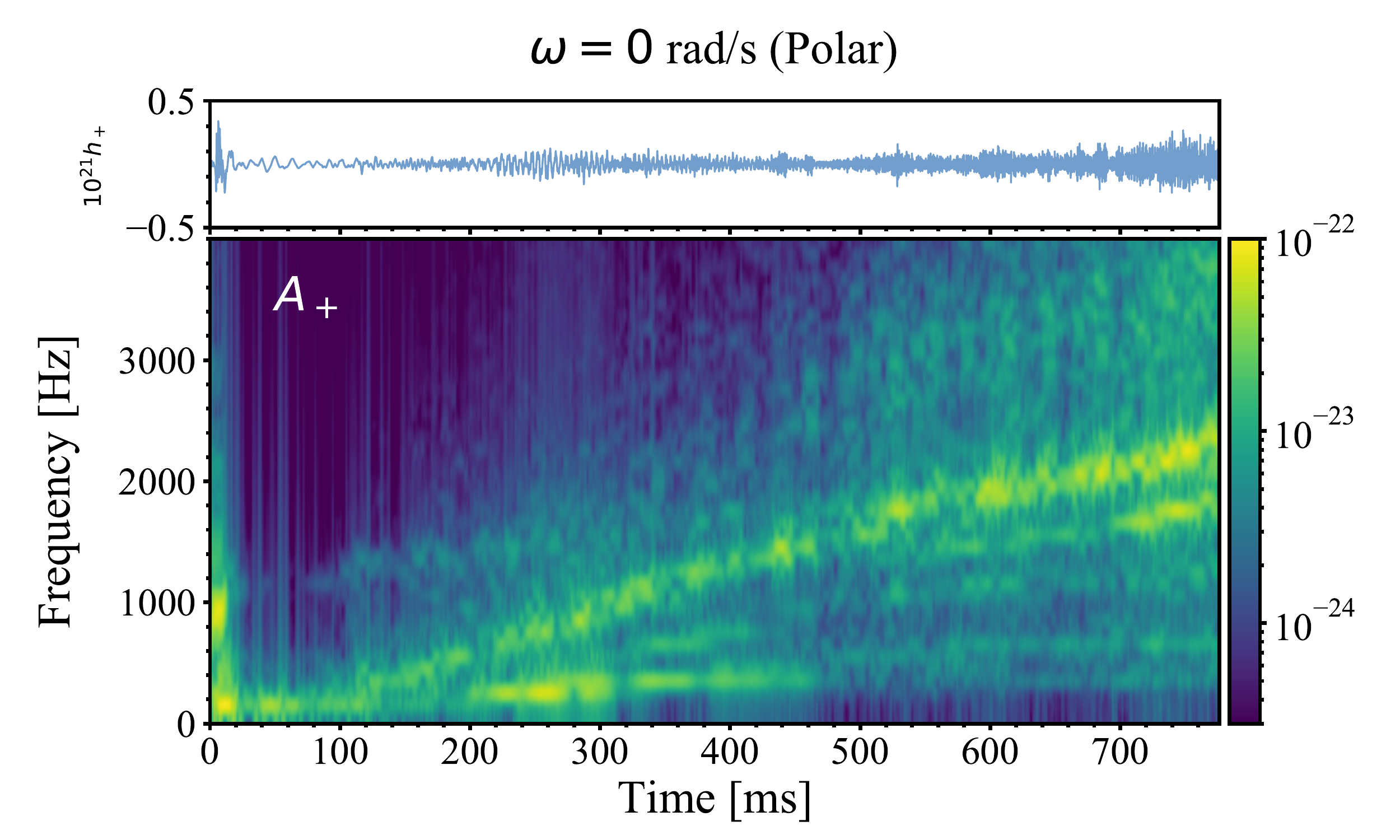}
	\plotone{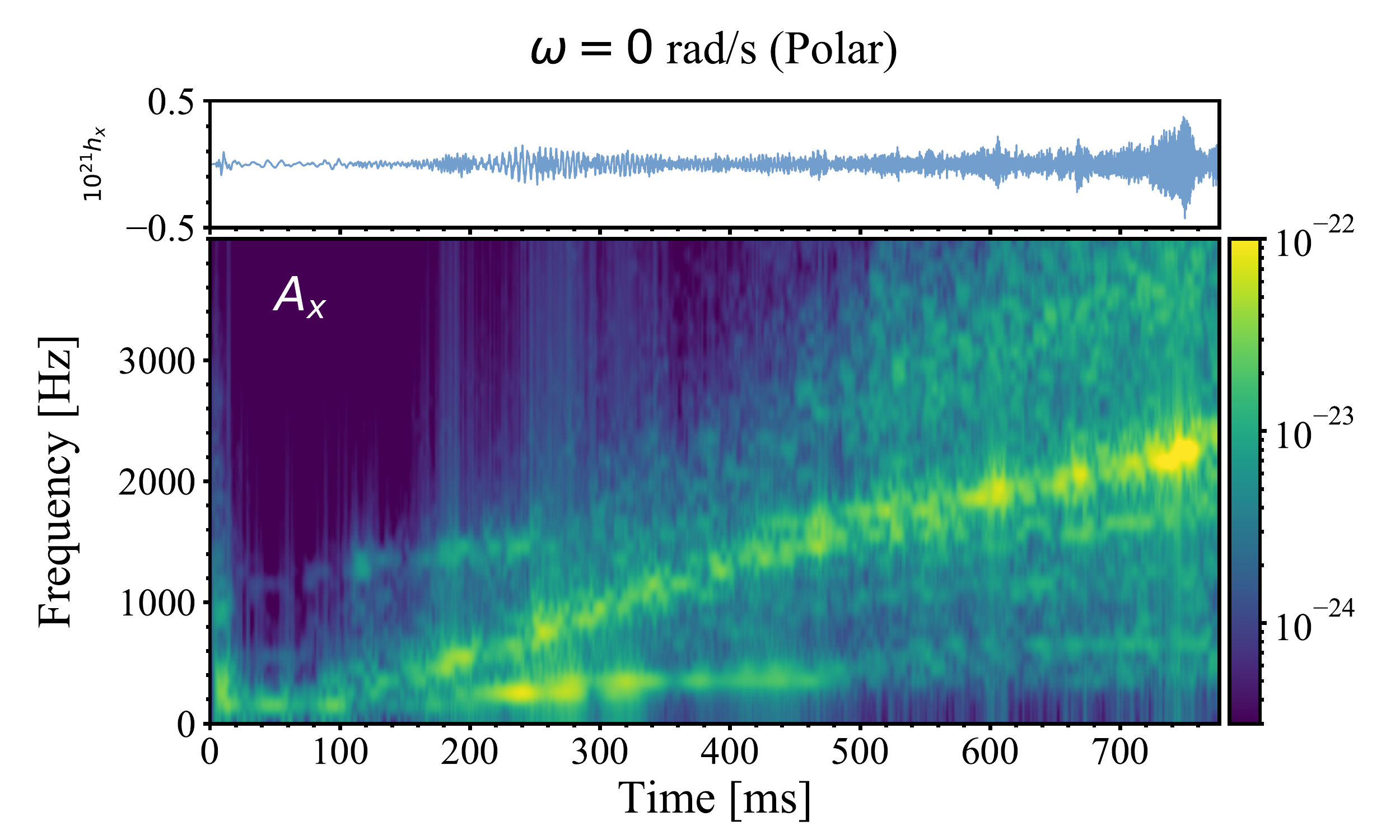}

	\plotone{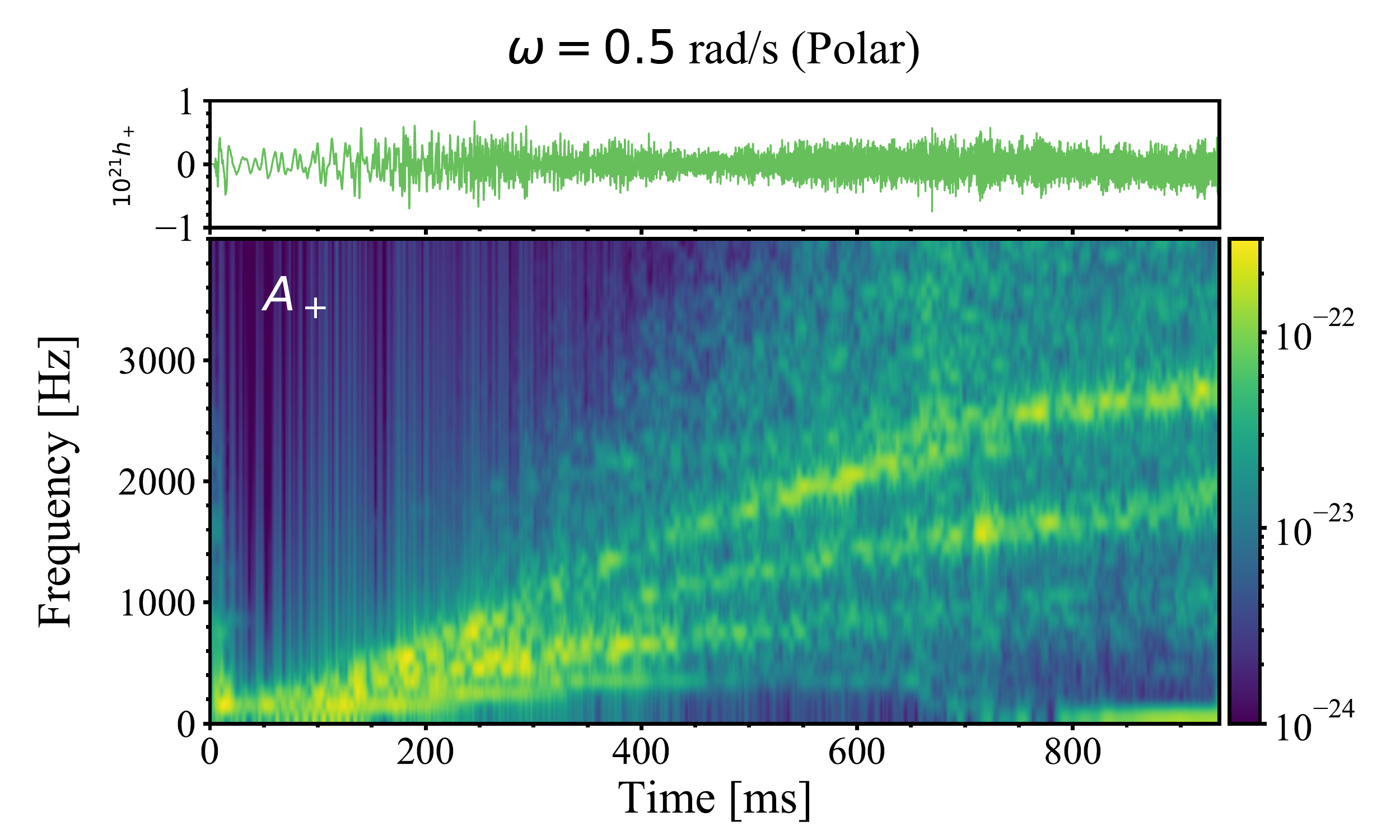}
	\plotone{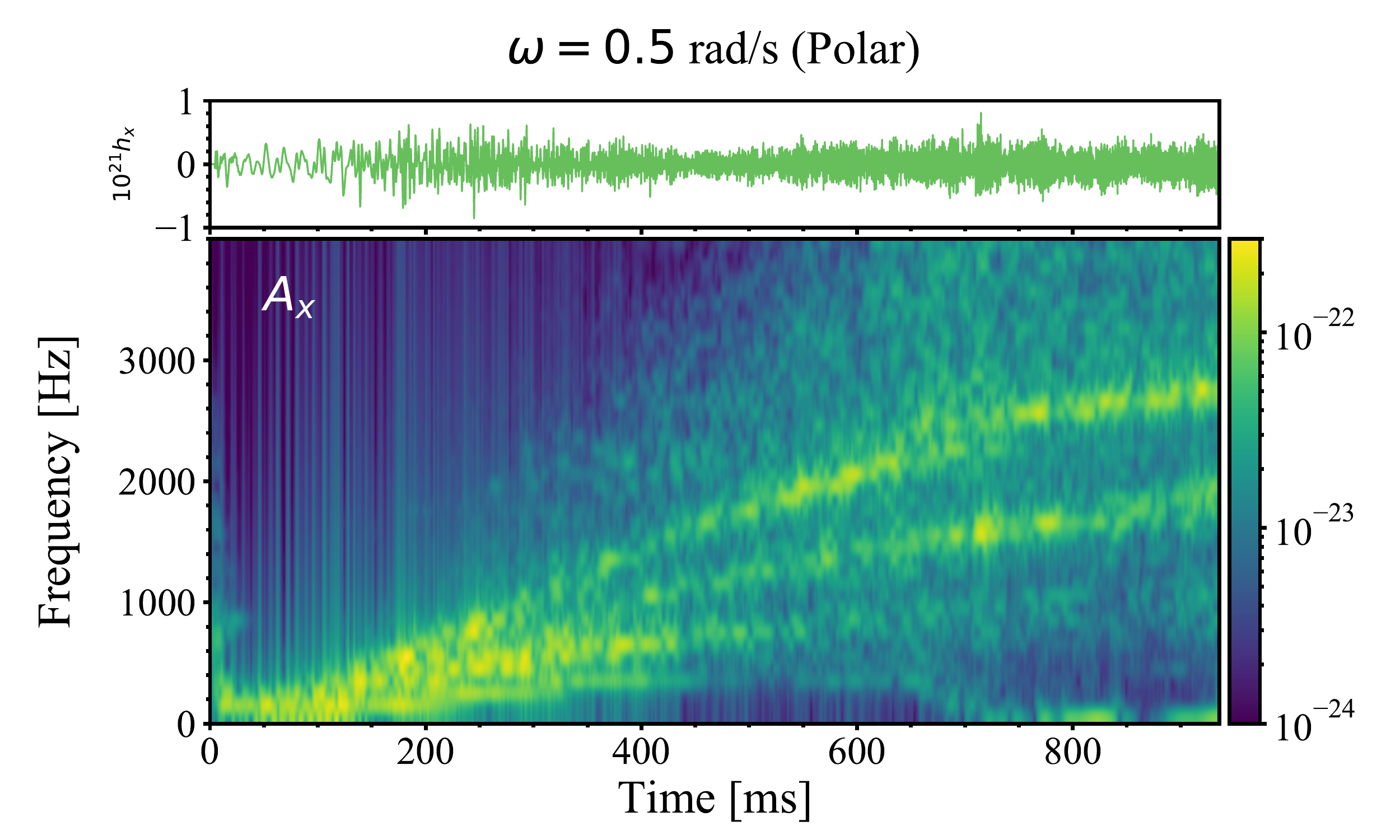}

	\plotone{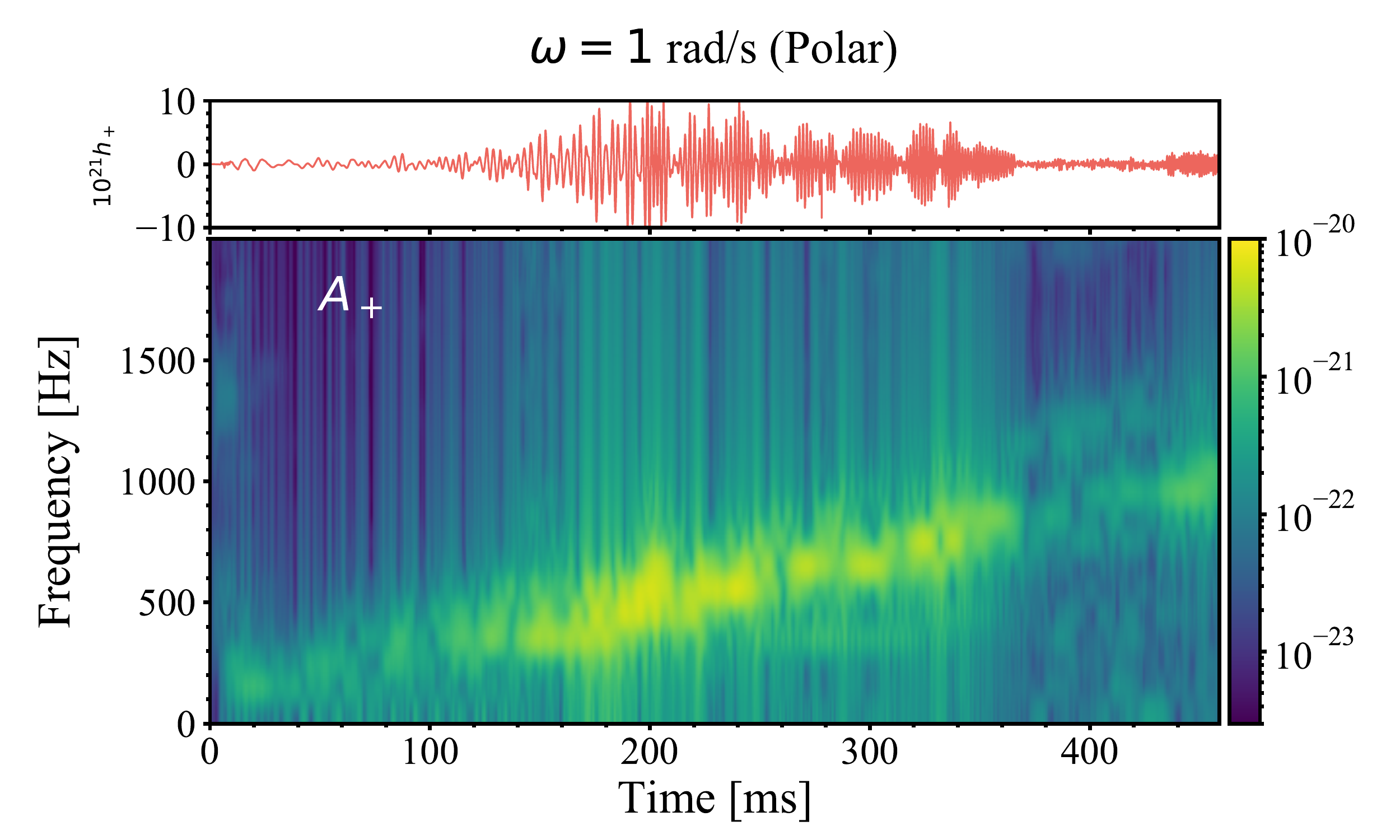}
	\plotone{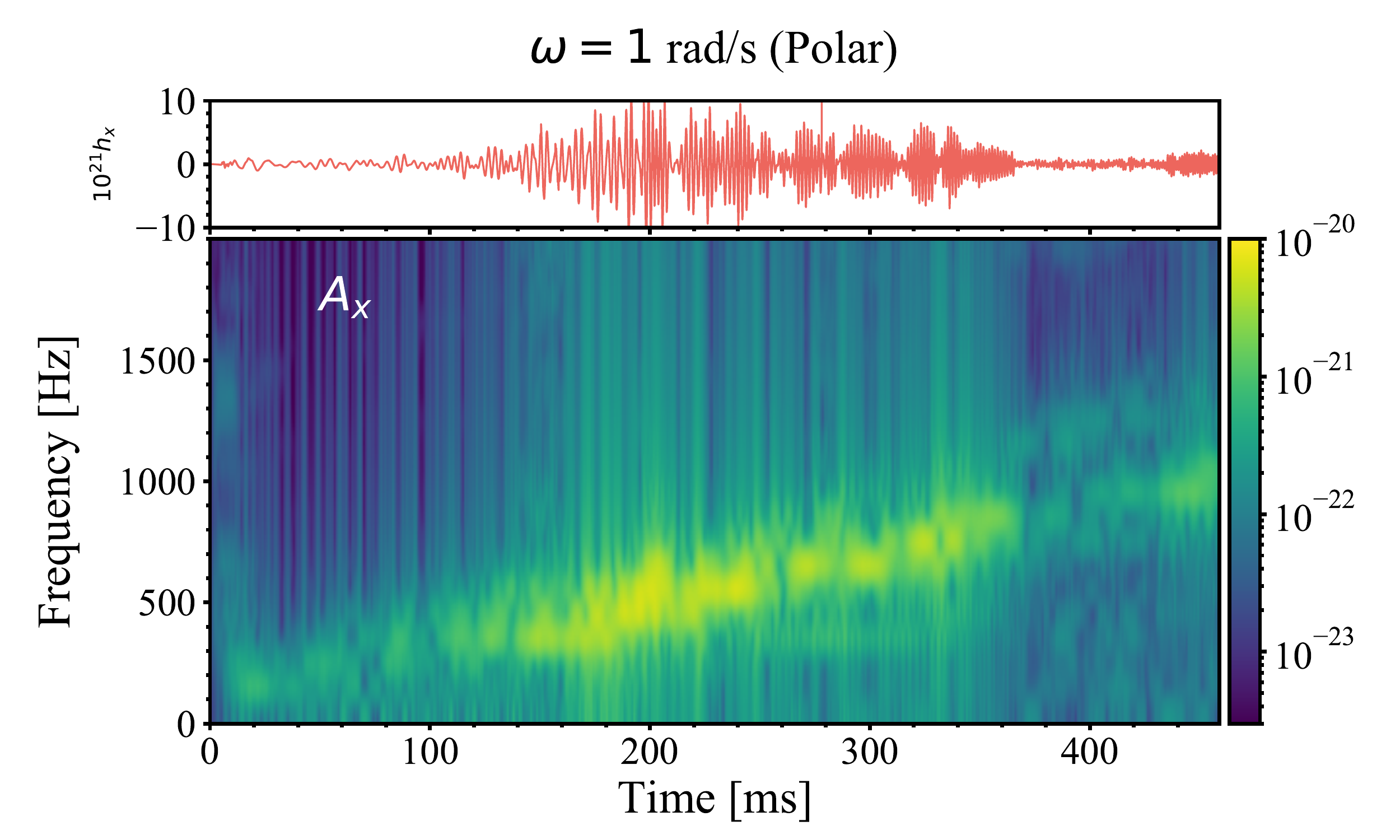}
	
	\caption{\label{fig_gw_polar}
	Similar to Figure~\ref{fig_gw_equator}, but as observed from the polar direction. }
	
\end{figure*}


We perform short-time Fourier transforms with a moving window of 10 ms to evaluate the GW spectrogram 
(Figures~\ref{fig_gw_equator} and \ref{fig_gw_polar}).
Overall, the NR model behaves very similar to its 2D counter part in \cite{2018ApJ...857...13P} and 
shows no difference between the plus mode and the cross mode in both polar and equatorial viewing angles.  
The PNS peak frequency can be seen in the extended yellow band in Figure~\ref{fig_gw_equator} and \ref{fig_gw_polar}.
which has been identified widely in various simulations 
\citep{2013ApJ...766...43M, 2013ApJ...779L..18C, 2016MNRAS.461L.112T, 2017MNRAS.468.2032A, 2019MNRAS.486.2238A}. 

The low-frequency components ($<200$~Hz) are expected to come from the SASI or SASI-excited modes inside the PNS
\citep{2016ApJ...829L..14K, 2018ApJ...865...81O, 2019ApJ...876L...9R, 2019MNRAS.486.2238A}. 
Although the strength of the SASI GW emissions is much lower than that of the PNS peak oscillations, the frequency window lies on the most 
sensitive band of the advanced LIGO, Virgo, and KAGRA. Therefore, the SASI signals could be useful to disentangle 
the degeneracy caused by microphysics and CCSN progenitor.    
A Fourier analysis of $l=1$ modes of our SASI component shown in Figure~\ref{fig_sasi} gives a low-frequency component at around 200~Hz, 
which is apparent in Figures~\ref{fig_gw_equator} and \ref{fig_gw_polar} at late times.
Additional features with frequencies lower than the PNS peak frequency in model SR and NR can be seen in Figure~\ref{fig_gw_equator} and \ref{fig_gw_polar}.
These features may came from higher-order modes of SASI or higher-order $g-/p-$ modes.

GW emission from rapidly rotating progenitors has been studied in several works, including  
\cite{2005ApJ...625L.119O, 2010A&A...514A..51S, 2014PhRvD..89d4011K, 2020MNRAS.tmpL..27S, 2020MNRAS.tmp.1245P}.
Under some extreme rotational conditions the so-called low $T/W$ instability \citep{2005ApJ...625L.119O} can develop, resulting in a bar-mode-like configuration of the rotating PNS. 
A strong GW signal from the one- or two-arm spiral waves from the PNS surface is observed in \cite{2020MNRAS.tmpL..27S}.
Although the $70 \Msun$ progenitor used in \cite{2020MNRAS.tmpL..27S} is different from ours, 
the FR model shows a similar component starting from $\sim$150~ms at $\sim$450~Hz. The strength of this GW signal then grows in time.
Note that the initial angular velocity of the iron core in \cite{2020MNRAS.tmpL..27S} is about $2$~rad~s$^{-1}$, 
which is twice that of our FR model.

\begin{figure*}
	\epsscale{1.0}
	\plotone{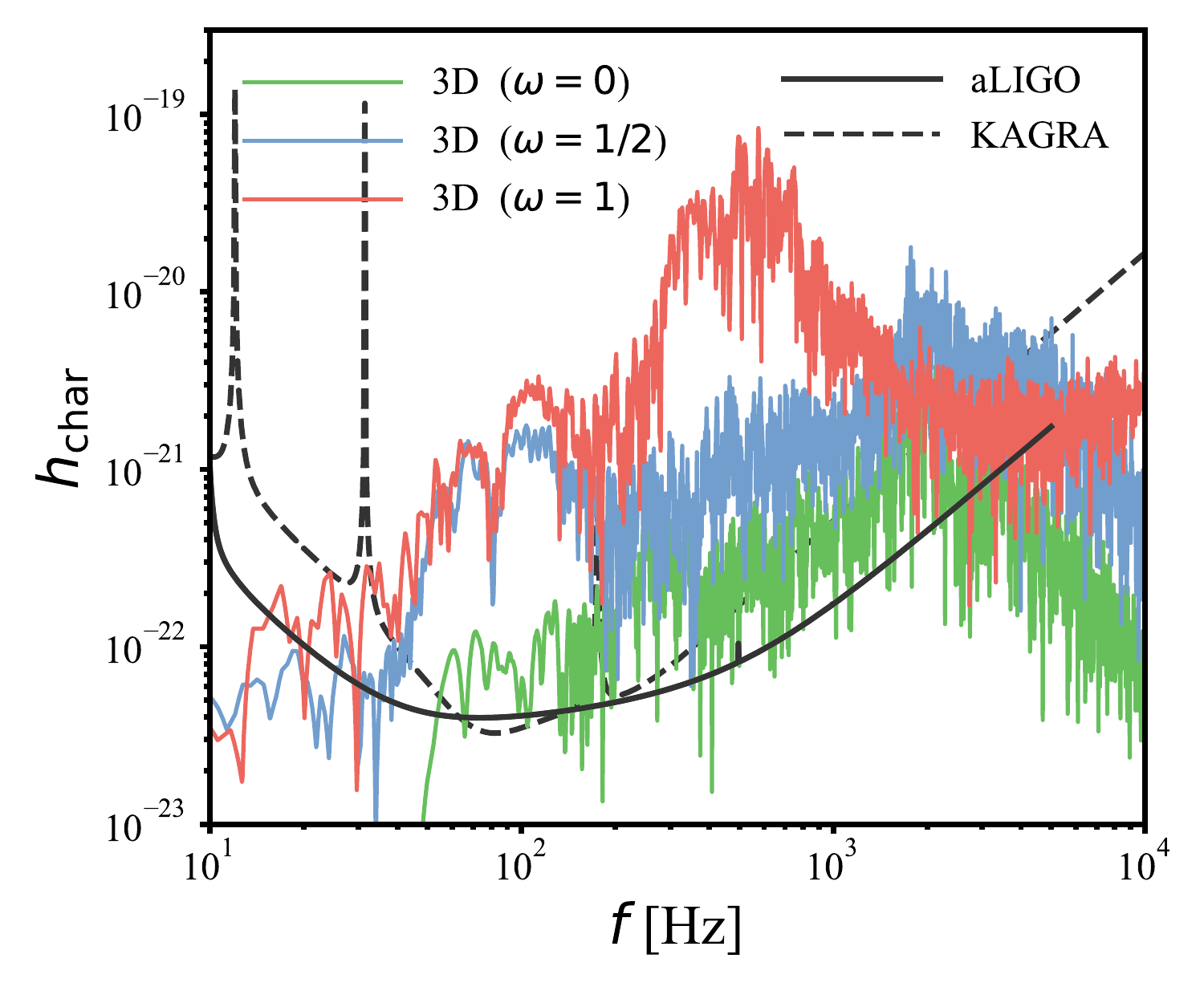}
	\caption{\label{fig_spectra}
	The characteristic GW amplitudes, $h_{char}$, for the three 3D models with different initial rotational speeds at 10~kpc.
	The solid (dashed) black line represents the designed sensitivity curve of the advanced-LIGO (KAGRA). }
\end{figure*}

Figure~\ref{fig_spectra} shows the dimensionless characteristic amplitudes ($h_{\rm char}$) as functions of frequency. 
The $h_{\rm char}$ is calculated based on the formula described in \cite{2009ApJ...707.1173M} and \cite{2018ApJ...857...13P}.
The sensitivity curves from the advanced LIGO and KAGRA \citep{2015CQGra..32a5014M} are plotted together in the black solid and dashed lines. 
Assuming a distance of 10~kpc, all three models can be detected by  advanced LIGO, Virgo, and KAGRA.
It is noticeable that rotation can enhance the strength of the GW amplitudes, but the frequency ranges are comparable.
In model FR, the peak frequency is lower than in model SR and NR. This is mainly due to an earlier termination time in model FR  
such that the PNS is less compact than in model SR and NR which are terminated right before BH formation.   
In addition, model SR and FR have higher low-frequency components at $\sim 100$~Hz, 
which might come from the eigenfrequency of the $m=1$ mode of the rotation \citep{2020MNRAS.tmpL..27S}.  
The peak frequency at BH formation is above $\sim 2000$~Hz, which is consistent with our previous 2D simulations in \cite{2018ApJ...857...13P},
suggesting the next generation GW detectors should improve the sensitivity in the kHz window in order to explore the physics around BH formation \citep{2016PhRvD..93d2002G, 2018ApJ...857...13P}.

\section{SUMMARY AND CONCLUSIONS} \label{sec_conclusion}

We have performed 3D core-collapse supernova simulations of the s40 progenitor from \cite{2007PhR...442..269W} with three different rotational speeds. 
The fast-rotating model explodes early at $\sim 250$~ms postbounce 
and emits strong gravitational wave signals 
at $400-800$~Hz.  
The non-rotating and slow-rotating models form BHs at 776~ms and 936~ms postbounce.
While the non-rotating model is a failed supernova, shock revival is observed in the slow-rotating model 
at about 180~ms prior to the BH formation.
This scenario is similar to what has been observed in \cite{2018MNRAS.477L..80K} and \cite{2018ApJ...852L..19C} 
but has a different time scale for BH formation and fallback accretion.    
Both exploding models have the diagnostic explosion energy higher than $6\times 10^{50}$~erg and continue growing in time. 
The bound remnant mass at the end of the simulation in the SR model is $3.46 \Msun$, 
which could explain the origin of the low mass BH ($2.6 \Msun$) in GW190814 \citep{2020ApJ...896L..44A}.

In addition to the time delay of BH formation in the slow-rotating model, 
we also find that SASI could help to redistribute angular momentum and could excite PNS rotation even in the non-rotating model as it was suggested in \cite{2007Natur.445...58B, 2018ApJ...865...81O}. 
In our FR model, explosion occurs without PNS collapse to a BH, leaving behind a rapidly rotating PNS with an induced millisecond rotation period. 
For the SR model, explosion is accompanied by subsequent collapse of the PNS to a BH, leaving behind a BH with spin parameter of $a= 0.52$, assuming specific angular momentum is conserved.

Our simulations show that the GW emission is detectable by the advanced LIGO, Virgo, and KAGRA 
if the sources are within 10~kpc. 
The bounce GW signal can be used to identify the angular momentum of the iron core, and 
the growth of the PNS oscillation throughout the very interesting regime of the EoS up to BH formation.
The peak GW frequencies in our models with BH formation are above $2000$~Hz, which are close to the threshold of current GW detectors.

\acknowledgments
KCP acknowledges valuable discussions with Michael Pajkos on the gravitational wave analysis.  
This work is supported by the Ministry of Science and Technology of Taiwan 
through grants MOST 107-2112-M-007-032-MY3, MOST 108-2811-M-007-562,
by the Center for Informatics and Computation in Astronomy (CICA) at National Tsing Hua University 
through a grant from the Ministry of Education of Taiwan;
by the European Research Council (ERC; FP7)
under ERC Advanced Grant Agreement N$^\circ$~321263~-~FISH;
by the Swiss National Science Foundation (SNF).
This article is based upon work from the “ChETEC” COST Action (CA16117), 
supported by COST (European Cooperation in Science and Technology).
SMC is supported by the U.S. Department of Energy, Office of Science, Office of Nuclear Physics, Early Career Research Program under Award Number DE-SC0015904. This material is based upon work supported by the U.S. Department of Energy, Office of Science, Office of Advanced Scientific Computing Research and Office of Nuclear Physics, Scientific Discovery through Advanced Computing (SciDAC) program under Award Number DE- SC0017955. This research was supported by the Exascale Computing Project (17-SC-20-SC), a collaborative effort of the U.S. Department of Energy Office of Science and the National Nuclear Security Administration.
{\tt FLASH} was in part developed by the DOE NNSA-ASC OASCR Flash Center 
at the University of Chicago.
The simulations and data analysis have been carried out at the CSCS ({\tt Piz-Daint}) under grant No.~s661 and s840,
on the {\tt Taiwania} supercomputer in the National Center for High-Performance Computing (NCHC) in Taiwan, 
and on the high performance computing center (HPCC) at Michigan State University.
Analysis and visualization of simulation data were completed using the analysis toolkit {\tt yt}
\citep{2011ApJS..192....9T}.
\software{FLASH \citep{2000ApJS..131..273F, 2008PhST..132a4046D}, yt \citep{2011ApJS..192....9T}, Matplotlib \citep{2007CSE.....9...90H}, NumPy \citep{2011CSE....13b..22V}, SciPy \citep{2019zndo...3533894V}}


\end{document}